\documentstyle[12pt,epsf]{article}
\begin{document}
\newcommand {\ee}{\end{equation}}
\newcommand {\bea}{\begin{eqnarray}}
\newcommand {\eea}{\end{eqnarray}}
\newcommand {\nn}{\nonumber \\}
\newcommand {\Tr}{{\rm Tr\,}}
\newcommand {\tr}{{\rm tr\,}}
\newcommand {\e}{{\rm e}}
\newcommand {\etal}{{\it et al.}}
\newcommand {\m}{\mu}
\newcommand {\n}{\nu}
\newcommand {\pl}{\partial}
\newcommand {\p} {\phi}
\newcommand {\vp}{\varphi}
\newcommand {\vpc}{\varphi_c}
\newcommand {\al}{\alpha}
\newcommand {\be}{\beta}
\newcommand {\ga}{\gamma}
\newcommand {\Ga}{\Gamma}
\newcommand {\x}{\xi}
\newcommand {\ka}{\kappa}
\newcommand {\la}{\lambda}
\newcommand {\La}{\Lambda}
\newcommand {\si}{\sigma}
\newcommand {\th}{\theta}
\newcommand {\Th}{\Theta}
\newcommand {\om}{\omega}
\newcommand {\Om}{\Omega}
\newcommand {\ep}{\epsilon}
\newcommand {\vep}{\varepsilon}
\newcommand {\na}{\nabla}
\newcommand {\del}  {\delta}
\newcommand {\Del}  {\Delta}
\newcommand {\mn}{{\mu\nu}}
\newcommand {\ls}   {{\lambda\sigma}}
\newcommand {\ab}   {{\alpha\beta}}
\newcommand {\half}{ {\frac{1}{2}} }
\newcommand {\fourth} {\frac{1}{4} }
\newcommand {\sixth} {\frac{1}{6} }
\newcommand {\sqg} {\sqrt{g}}
\newcommand {\fg}  {\sqrt[4]{g}}
\newcommand {\invfg}  {\frac{1}{\sqrt[4]{g}}}
\newcommand {\sqZ} {\sqrt{Z}}
\newcommand {\gbar}{\bar{g}}
\newcommand {\sqk} {\sqrt{\kappa}}
\newcommand {\sqt} {\sqrt{t}}
\newcommand {\reg} {\frac{1}{\epsilon}}
\newcommand {\fpisq} {(4\pi)^2}
\newcommand {\Lcal}{{\cal L}}
\newcommand {\Ocal}{{\cal O}}
\newcommand {\Dcal}{{\cal D}}
\newcommand {\Ncal}{{\cal N}}
\newcommand {\Mcal}{{\cal M}}
\newcommand {\scal}{{\cal s}}
\newcommand {\Dvec}{{\hat D}}   
\newcommand {\dvec}{{\vec d}}
\newcommand {\Evec}{{\vec E}}
\newcommand {\Hvec}{{\vec H}}
\newcommand {\Vvec}{{\vec V}}
\newcommand {\Btil}{{\tilde B}}
\newcommand {\ctil}{{\tilde c}}
\newcommand {\Ftil}{{\tilde F}}
\newcommand {\Stil}{{\tilde S}}
\newcommand {\Ztil}{{\tilde Z}}
\newcommand {\altil}{{\tilde \alpha}}
\newcommand {\betil}{{\tilde \beta}}
\newcommand {\latil}{{\tilde \lambda}}
\newcommand {\ptil}{{\tilde \phi}}
\newcommand {\Ptil}{{\tilde \Phi}}
\newcommand {\natil} {{\tilde \nabla}}
\newcommand {\ttil} {{\tilde t}}
\newcommand {\Shat}{{\hat S}}
\newcommand {\shat}{{\hat s}}
\newcommand {\Dhat}{{\hat D}}   
\newcommand {\Vhat}{{\hat V}}   
\newcommand {\xhat}{{\hat x}}
\newcommand {\Zhat}{{\hat Z}}
\newcommand {\Gahat}{{\hat \Gamma}}
\newcommand {\nah} {{\hat \nabla}}
\newcommand {\gh}  {{\hat g}}
\newcommand {\labar}{{\bar \lambda}}
\newcommand {\cbar}{{\bar c}}
\newcommand {\bbar}{{\bar b}}
\newcommand {\Bbar}{{\bar B}}
\newcommand {\psibar}{{\bar \psi}}
\newcommand {\chibar}{{\bar \chi}}
\newcommand {\bbartil}{{\tilde {\bar b}}}
\newcommand {\intfx} {{\int d^4x}}
\newcommand {\inttx} {{\int d^2x}}
\newcommand {\change} {\leftrightarrow}
\newcommand {\ra} {\rightarrow}
\newcommand {\larrow} {\leftarrow}
\newcommand {\ul}   {\underline}
\newcommand {\pr}   {{\quad .}}
\newcommand {\com}  {{\quad ,}}
\newcommand {\q}    {\quad}
\newcommand {\qq}   {\quad\quad}
\newcommand {\qqq}   {\quad\quad\quad}
\newcommand {\qqqq}   {\quad\quad\quad\quad}
\newcommand {\qqqqq}   {\quad\quad\quad\quad\quad}
\newcommand {\qqqqqq}   {\quad\quad\quad\quad\quad\quad}
\newcommand {\qqqqqqq}   {\quad\quad\quad\quad\quad\quad\quad}
\newcommand {\lb}    {\linebreak}
\newcommand {\nl}    {\newline}

\newcommand {\vs}[1]  { \vspace*{#1 cm} }

\newcommand {\MPL}  {Mod.Phys.Lett.}
\newcommand {\NP}   {Nucl.Phys.}
\newcommand {\PL}   {Phys.Lett.}
\newcommand {\PR}   {Phys.Rev.}
\newcommand {\PRL}   {Phys.Rev.Lett.}
\newcommand {\CMP}  {Commun.Math.Phys.}
\newcommand {\JMP}  {Jour.Math.Phys.}
\newcommand {\AP}   {Ann.of Phys.}
\newcommand {\PTP}  {Prog.Theor.Phys.}
\newcommand {\NC}   {Nuovo Cim.}
\newcommand {\CQG}  {Class.Quantum.Grav.}


\font\smallr=cmr5
\def\ocirc#1{#1^{^{{\hbox{\smallr\llap{o}}}}}}
\def\ogamma{\ocirc{\gamma}{}}
\def\oM{{\buildrel {\hbox{\smallr{o}}} \over M}}
\def\osigma{\ocirc{\sigma}{}}

\def\overleftrightarrow#1{\vbox{\ialign{##\crcr
 $\leftrightarrow$\crcr\noalign{\kern-1pt\nointerlineskip}
 $\hfil\displaystyle{#1}\hfil$\crcr}}}
\def\overnab{{\overleftrightarrow\nabslash}}

\def\va{{a}}
\def\vb{{b}}
\def\vc{{c}}
\def\tilpsi{{\tilde\psi}}
\def\tbpsi{{\tilde{\bar\psi}}}

\def\Dslash{{}\hbox{\hskip2pt\vtop
 {\baselineskip23pt\hbox{}\vskip-24pt\hbox{/}}
 \hskip-11.5pt $D$}}
\def\nabslash{{}\hbox{\hskip2pt\vtop
 {\baselineskip23pt\hbox{}\vskip-24pt\hbox{/}}
 \hskip-11.5pt $\nabla$}}
\def\xislash{{}\hbox{\hskip2pt\vtop
 {\baselineskip23pt\hbox{}\vskip-24pt\hbox{/}}
 \hskip-11.5pt $\xi$}}
\def\leftnabla{{\overleftarrow\nabla}}

\def\delL{{\delta_{LL}}}
\def\delG{{\delta_{G}}}
\def\delc{{\delta_{cov}}}

\newcommand {\sqxx}  {\sqrt {x^2+1}}   
\newcommand {\gago}  {\gamma_5}
\newcommand {\Ktil}  {{\tilde K}}
\newcommand {\Ltil}  {{\tilde L}}
\newcommand {\Qtil}  {{\tilde Q}}
\newcommand {\Rtil}  {{\tilde R}}
\newcommand {\Kbar}  {{\bar K}}
\newcommand {\Lbar}  {{\bar L}}
\newcommand {\Qbar}  {{\bar Q}}
\newcommand {\Pp}  {P_+}
\newcommand {\Pm}  {P_-}
\newcommand {\GfMp}  {G^{5M}_+}
\newcommand {\GfMpm}  {G^{5M'}_-}
\newcommand {\GfMm}  {G^{5M}_-}
\newcommand {\Omp}  {\Omega_+}    
\newcommand {\Omm}  {\Omega_-}
\def\Aslash{{}\hbox{\hskip2pt\vtop
 {\baselineskip23pt\hbox{}\vskip-24pt\hbox{/}}
 \hskip-11.5pt $A$}}
\def\Rslash{{}\hbox{\hskip2pt\vtop
 {\baselineskip23pt\hbox{}\vskip-24pt\hbox{/}}
 \hskip-11.5pt $R$}}
\def\kslash{
{}\hbox       {\hskip2pt\vtop
                   {\baselineskip23pt\hbox{}\vskip-24pt\hbox{/}}
               \hskip-8.5pt $k$}
           }    
\def\qslash{
{}\hbox       {\hskip2pt\vtop
                   {\baselineskip23pt\hbox{}\vskip-24pt\hbox{/}}
               \hskip-8.5pt $q$}
           }    
\def\dslash{
{}\hbox       {\hskip2pt\vtop
                   {\baselineskip23pt\hbox{}\vskip-24pt\hbox{/}}
               \hskip-8.5pt $\partial$}
           }    
\def\dbslash{{}\hbox{\hskip2pt\vtop
 {\baselineskip23pt\hbox{}\vskip-24pt\hbox{$\backslash$}}
 \hskip-11.5pt $\partial$}}
\def\Kbslash{{}\hbox{\hskip2pt\vtop
 {\baselineskip23pt\hbox{}\vskip-24pt\hbox{$\backslash$}}
 \hskip-11.5pt $K$}}
\def\Ktilbslash{{}\hbox{\hskip2pt\vtop
 {\baselineskip23pt\hbox{}\vskip-24pt\hbox{$\backslash$}}
 \hskip-11.5pt ${\tilde K}$}}
\def\Ltilbslash{{}\hbox{\hskip2pt\vtop
 {\baselineskip23pt\hbox{}\vskip-24pt\hbox{$\backslash$}}
 \hskip-11.5pt ${\tilde L}$}}
\def\Qtilbslash{{}\hbox{\hskip2pt\vtop
 {\baselineskip23pt\hbox{}\vskip-24pt\hbox{$\backslash$}}
 \hskip-11.5pt ${\tilde Q}$}}
\def\Rtilbslash{{}\hbox{\hskip2pt\vtop
 {\baselineskip23pt\hbox{}\vskip-24pt\hbox{$\backslash$}}
 \hskip-11.5pt ${\tilde R}$}}
\def\Kbarbslash{{}\hbox{\hskip2pt\vtop
 {\baselineskip23pt\hbox{}\vskip-24pt\hbox{$\backslash$}}
 \hskip-11.5pt ${\bar K}$}}
\def\Lbarbslash{{}\hbox{\hskip2pt\vtop
 {\baselineskip23pt\hbox{}\vskip-24pt\hbox{$\backslash$}}
 \hskip-11.5pt ${\bar L}$}}
\def\Rbarbslash{{}\hbox{\hskip2pt\vtop
 {\baselineskip23pt\hbox{}\vskip-24pt\hbox{$\backslash$}}
 \hskip-11.5pt ${\bar R}$}}
\def\Qbarbslash{{}\hbox{\hskip2pt\vtop
 {\baselineskip23pt\hbox{}\vskip-24pt\hbox{$\backslash$}}
 \hskip-11.5pt ${\bar Q}$}}
\def\Acalbslash{{}\hbox{\hskip2pt\vtop
 {\baselineskip23pt\hbox{}\vskip-24pt\hbox{$\backslash$}}
 \hskip-11.5pt ${\cal A}$}}

\begin{flushright}
August 1999\\
hep-th/9908156 \\
US-99-02
\end{flushright}

\vspace{0.5cm}

\begin{center}
{\Large\bf 
New Regularization Using Domain Wall }

\vspace{1.5cm}
{\large Shoichi ICHINOSE
          \footnote{
E-mail address:\ ichinose@u-shizuoka-ken.ac.jp
                  }
}
\vspace{1cm}

{\large 
Department of Physics, University of Shizuoka,
Yada 52-1, Shizuoka 422-8526, Japan          
}

\end{center}
\vfill

{\large Abstract}\nl
We present a new regularization method, for 
d dim (Euclidean) quantum field theories in the continuum formalism, 
based on the domain wall configuration in (1+d) dim space-time.
It is inspired by the recent progress in the 
chiral fermions on lattice. 
The wall "height" is given by $1/M$, where $M$ is 
a regularization mass parameter and appears as a
(1+d) dim Dirac fermion mass. The present approach gives 
a {\it thermodynamic view} to the domain wall or the overlap formalism
in the lattice field theory. 
We will show qualitative correspondence between the present 
continuum results and those of lattice.
The extra dimension 
is regarded as the (inverse) {\it temperature} $t$.
The domains are defined by the {\it directions} of
the "system evolvement", not by
the sign of $M$ as in the original overlap formalism.
Physically the parameter  $M$ controls both the
chirality selection and the dimensional reduction to
$d$ dimension (domain wall formation). 
From the point of regularization, the limit $Mt\ra 0$
regularize the infra-red behaviour whereas the condition
on the momentum ($k^\m$) integral, $|k^\m|\leq M$, regularize
the ultra-violet behaviour.

To check the new regularization works correctly, we take 
the 4 dim QED 
and 2 dim chiral gauge theory as examples.
Especially the consistent and covariant
anomalies are correctly obtained. 
The choice of solutions of the higher dim
Dirac equation characterizes the two anomalies.
The projective properties of the positive and negative energy free solutions
are exploited in calculation.
Some integral functions, the incomplete gamma functions
and the generalized hypergeometric functions 
characteristically appear in this new regularization
procedure.

\vspace{0.5cm}

PACS NO:\ 11.30.Rd,11.25.Db,05.70.-a,11.10.Kk,11.10.Wx \nl
Key Words:\ Chiral Fermion, Domain Wall, Overlap Formalism,
Regularization, Heat-kernel, Chiral Anomaly

\newpage

\section{Introduction}
\q Regularizing quantum theories respecting the chirality has been a long-lasting problem
both in the discrete and in the continuum field theories. 
The difficulty originates
from the fact that the chiral symmetry is a symmetry 
strongly bound to the space-time dimension 
and is related to the discrete symmetry of parity
and to the global features of the space-time topology.
Non-continuous property is usually difficult to regularize. 
Ordinary regularizations, such as the dimensional regularization, 
often hinder controlling the chirality.
The symmetry should be compared
with others such as the gauge symmetry of the internal space
and the Lorentz symmetry of the space-time.
In the lattice field theory, the difficulty appears as the doubling
problem of fermions\cite{Wil75} (see a text, say, \cite{Creu83})
and as the Nielsen-Ninomiya no-go theorem
\cite{NN81}. 
The recent very attractive progress in the lattice chiral fermion
tells us the domain wall configuration in one dimension higher
space(-time) serves as a good regularization
, at least, as far as vector theories are concerned
\footnote{
We do not touch on the chiral gauge theories except
some argument in the second paragraph of Sect.8.
}
\cite{Kap92,Jan92}. 
It was formulated as the overlap formalism\cite{NN94,NN95}
and was further examined by \cite{DS95PL,DS95NP,DS97PL}.
The corresponding lattice models were analyzed by 
\cite{Sha93,CH94,Creu95,FS95,Vra98}.
The numerical data also look to support its validity\cite{Col99}.
Most recently 
the overlap Dirac operator by Neuberger\cite{Neub98},
which satisfies Ginsparg-Wilson relation\cite{GW82}, and
L{\"{u}}scher's chiral symmetry on lattice\cite{Lus98}
makes the present direction more and more attractive.

\q The present motivation is to find a counter-part, in the continuum,
of the above regularization on lattice.
Through the analysis we expect to clarify the essence of
the regularization mechanism more transparently than on lattice. 
We see some advanced points over the  
ordinary regularizations in the continuum field theories. 
The main goal is to develop a new
feasible regularization, in the continuum formalism,  
which is compatible with the chiral symmetry.

\q The overlap formalism has been 
newly formulated using the heat-kernel\cite{SI98}. 
The heat-kernel formalism is most efficiently expressed in the
coordinate space\cite{II98}, and which enables us to do comparison
with the lattice formalism. We will often
compare the present results with those obtained by 
the lattice domain wall approach.
The present formalism is based on three key points:
\begin{enumerate} 
\item
We utilize the characteristic relation of heat and temperature, 
that is, heat propagates
from the high temperature to the low temperature 
(the second law of thermodynamics). 
In the system which obey the heat equation, there exists
a {\it fixed direction} in the system evolvement. 
We regard the heat equation
for the spinor system, after the Wick rotation,  
as the Dirac equation in one dimension higher space-time. And
we consider the (1+4 dim classical) configuration 
which has a fixed direction in time.
This setting is suitable for regularizing the dynamics
(in 4 dim Euclidean space) with control of the chirality. 
\item 
Anti-commutativity between the system operator $\Dhat$
and the chiral matrix, that is, $\gago\Dhat+\Dhat\gago=0$
plays the crucial role 
to separate the whole configuration
into two parts (we will call them "(+)-domain" and "(-)-domain") 
which are related by the sign change of the "time"-axis. 
This is contrasted with
the original formulation of the overlap where
the difference of two vacua, one is constructed from
the (+) sign regularization (1+4 dim fermion) mass and the other from
the (-) sign,
distinguishes the two domains.
\item
Taking the small momentum region compared to the regularization
mass scale $M$ regularize the ultra-violet divergences 
and, at the same time, controls the chirality.
\end{enumerate}

\q 
In this paper we further examine the new approach and strengthen its basis
for the establishment.  
In order to show that the present approach is 
a new regularization for general field theories, 
the regularization mechanism is systematically presented.
Three kinds of wall configurations appear depending on the choice
of propagators (regularization).  
In the regularization of momentum-integral,  
some characteristic functions ( sine integral
functions, incomplete gamma functions, etc) appear. 
Because the presented perturbation calculation is not so familiar,
the description of calculation is rather explanatory so that
readers can follow them.
The main points in the present analysis are as follows.
\begin{enumerate}
\item
The extra axis is interpreted as the (inverse) temperature.
This formalism gives a thermodynamic view on the domain wall algorithm.
As the extra axis, it should be a half line ( not a line ) like the
temperature. 
\item
The characteristic condition of the present regularization
:\ $|k^\m|\leq |M|\ll 1/t$, is naturally obtained
and its role is closely examined.
\item
The reason why the ``overlap'' of $|+>$ and $|->$ should
be taken in the anomaly calculation is manifest. 
The ``overlap''
in the partition function corresponds to a ``difference''
in the effective action.
\item
This new regularization is applied to 4 dim Euclidean QED
and 2 dim chiral gauge theory. Especially, in the latter model, 
both consistent and covariant anomalies appear depending on
the choice of solutions in the 1+4 dim Dirac equation.
It is a new characterization of the two anomalies.
%
\end{enumerate}

\q In Sec.2 the heat-kernel method is reviewed taking 4 dim Euclidean
QED as an example. In Sec.3 the new regularization formalism 
of domain wall approach is explained. We apply it, in Sec.4, to the model
of Sec.2 and reproduce the result using the domain wall regularization.
In Sec.5 the ultraviolet regularization of the momentum integral 
is closely explained. Some characteristic functions appear.
The new regularization is applied to 2 dim chiral gauge theory in Sec.6.
The consistent and covariant anomalies are newly characterized. In Sec.7
we consider the case of massive fermion. Finally we conclude in Sec.8.
Three appendices are in order. App.A describes the present notation.
Some useful integral formulae are listed in App.B. Projective
properties of free solutions ( positive and negative energy ) 
of 1+4 dim Dirac equation are displayed in App.C. 

\section{QED in the Heat-Kernel Approach}
We review 4 dim Euclidean QED using the heat-kernel approach
and fix the present notation.
The results will be compared with the domain wall approach in
the following sections.
In 1951 Schwinger \cite{Sch51} did the heat-kernel analysis of 
the (1+3 dim) QED
and succeeded in calculating the radiative corrections in the
covariant way. Physically some interesting
phenomenon of the vacuum polarization in the strong magnetic
field was pointed out.

We consider the massless Euclidean case and focus on its
anomaly aspect. 
The lagrangian is given by
\begin{eqnarray}
\Lcal=i \psibar\Dslash\psi\com\q
\Dslash=\ga_\m(\pl_\m+i\e A_\m)\com\q
(i\Dslash)^\dag=i\Dslash\pr
\label{QED1}
\end{eqnarray}
Our convention of the gamma matrices is given in
App.A. 
The (1-loop) effective action is usually evaluated as
\begin{eqnarray}
\ln\,Z[A]=\ln\,
\int\Dcal\psibar\Dcal\psi\e^{-\intfx\Lcal}
=\Tr\ln i\Dslash
=\half\Tr\ln (-\Dslash^2)         \nn
=-\half \Tr\int_0^\infty\frac{\e^{-t\Dvec}}{t}dt+\mbox{const}\com\q
\Dvec=-\Dslash^2 \com
\label{QED2}
\end{eqnarray}
where $\Dvec$ is a quadratic differential operator
and has positive (semi)definite eigenvalues, hence
the $t$-integral converges well.  The heat kernel is
introduced as
\begin{eqnarray}
G(x,y;t)\equiv <x|e^{-t\Dvec}|y>\com                   \nn
(\frac{\pl}{\pl t}+\Dvec)G(x,y;t)=0 \com \q
\lim_{t\ra +0}G(x,y;t)=\del^4(x-y)\com            \nn
\ln\,Z[A]=-\half\int_0^\infty\frac{1}{t} \Tr G(x,y;t) dt+\mbox{const}\com
\label{QED3}
\end{eqnarray}
where $<x|$ and $|y>$ 
are the x-representation of $\Dhat$ (Dirac's bra-
and ket-vectors respectively) which can be well-defined
by the complete set of eigen-functions of $\Dhat$:\ 
$\{ f_n(x);n=0,1,\cdots |\ \Dhat f_n(x)=\la_nf_n(x)\}$.
The boundary condition equation in (\ref{QED3}) shows
the heat kernel regularization. The heat equation is
usually solved in perturbation around the free solution.
\begin{eqnarray}
\Dhat=-\pl^2-\Vhat\com\q
(\frac{\pl}{\pl t}-\pl^2)G(x,y;t) = \Vhat G(x,y;t)\com\nn
\Vhat=2ieA_\m\pl_\m-e^2A^2+ie\pl_\m A_\m
+\frac{ie}{4}[\ga_\m,\ga_\n]F_\mn
\com            \label{QED4}
\end{eqnarray}
where $F_\mn=\pl_\m A_\n-\pl_\n A_\m$.
For the anomaly calculation in this model, the relevant order is
$O(A^2)$. 
In terms of the free solution $G_0(x;t)$ and the propagator
$S(x;t)$ defined by
\begin{eqnarray}
(\frac{\pl}{\pl t}-\pl^2)G_0(x;t) = 0\com\q
\lim_{t\ra +0}G_0(x;t)=\del^4(x)I_4\com
\nn
(\frac{\pl}{\pl t}-\pl^2)S(x;t) = \del^4(x)\del(t)I_4\com
          \label{QED4.5}
\end{eqnarray}
the formal solution $G(x,y;t)$ is given as
\begin{eqnarray}
G(x,y;t)=G_0(x-y;t)+\int d^4z\int^\infty_{-\infty}ds~
S(x-z;t-s){\hat V}(z)G(z,y;s)               \com\nn
G_0(x-y;t)
=\frac{1}{(4\pi t)^2}e^{-\frac{(x-y)^2}{4t}}I_4\com\q
S(x-y;t)=\th(t)G_0(x-y;t)\com
\label{QED4b}
\end{eqnarray}
where $I_4$ is the 4 by 4 unit matrix. 
From these we obtain the boundary condition on $G(x,y;t)$.
\begin{eqnarray}
\lim_{t\ra +0}G(x,y;t)=\del^4(x-y)I_4\pr
          \label{QED4bb}
\end{eqnarray}
We note here the following things.
\begin{enumerate}
\item
The propagator has the form:\ the theta function $\th(t)$\ 
$\times$ free solution $G_0(x;t)$.
\item
The factor $\th(t)$
in $S(x-y;t)$ guarantees that the system evolves in the 
{\it forward direction} 
in the proper time $t$.
\end{enumerate}
These properties are characteristic
of the heat propagation and will be utilized in the
following sections. The analogy to the heat is the central idea of 
this paper\cite{SI98}. 

\q Under the infinitesimal chiral U(1) transformation:  
\begin{eqnarray}
\del_\al\psi=i\gago\al(x)\psi\com\q
\del_\al\psibar=\psibar i\gago\al(x)
\com            \label{QED5}
\end{eqnarray}
the lagrangian (\ref{QED1}) transforms as
\begin{eqnarray}
\del_\al\Lcal=-\pl_\m\al\psibar\ga_\m\gago\psi
\pr            \label{QED6}
\end{eqnarray}
Under the infinitesimal Weyl transformation
\begin{eqnarray}
\del_\om\psi=\om(x)\psi\com\q
\del_\om\psibar=\psibar \om(x)
\com            \label{QED7}
\end{eqnarray}
it changes as
\begin{eqnarray}
\del_\om\Lcal=i\pl_\m\om\psibar\ga_\m\psi+2\om\cdot\Lcal
\pr            \label{QED8}
\end{eqnarray}
The "naive" Ward-Takahashi(WT) identities for the chiral and Weyl transformations
are derived from (\ref{QED6}) and (\ref{QED8}). The deviation
from the "naive" WT identities, which originates from the measure change\cite{KF79}, 
is identified as the anomaly.
They are given by
\begin{eqnarray}
J_{ABJ}\equiv\left|\frac{\pl (\psi+\del_\al\psi,\psibar+\del_\al\psibar)}{\pl (\psi,\psibar)}\right|\com \nn 
\del_\al\,\ln\, J_{ABJ}
=\Tr\,i\al(x)\gago\,(\del^4(x-y)+{\bar {\del^4}}(x-y)\,)\ ,       \nn
\half\frac{\del}{\del\al(x)}\ln J_{ABJ}|_{\al=0}
=i\lim_{t\ra +0}\intfx\,\tr \gago G(x,x;t)
=\intfx\frac{e^2}{32\pi^2}iF_\mn\Ftil_\mn	   
\ ,            \label{QED9}
\end{eqnarray}
where $\Ftil_\mn=\ep_{\mn\ls}F_\ls$ , for the Adler-Bell-Jackiw (chiral U(1)) anomaly and
\begin{eqnarray}
J_W\equiv\left|\frac{\pl (\psi+\del_\om\psi,\psibar+\del_\om\psibar)}{\pl (\psi,\psibar)}\right|\com \nn
\del_\om\,\ln\, J_W
=\Tr\,\om(x)\,(\del^4(x-y)+{\bar {\del^4}}(x-y)\,)   \ ,\nn
\half\frac{\del}{\del\om(x)}\ln J|_{\om=0}
=\lim_{t\ra +0}\intfx\,\tr G(x,x;t)
=\half\intfx\,\be(e)F_\mn F_\mn	\nn
\be(e)=\frac{e^2}{12\pi^2}   
\com            \label{QED10}
\end{eqnarray}
for the Weyl anomaly. 
In (\ref{QED9}) and (\ref{QED10}), the delta functions
are regularized using the relation (\ref{QED4bb}). 
$\be(e)$ is the 1-loop $\be$-function
of QED. A useful formula for the calculation of $G(x,x;t=+0)$
in (\ref{QED9}) and (\ref{QED10}) is given in \cite{II96}.
(For the Yang-Mills theory, $\be$-function was obtained
in this way in \cite{KF93}.)

\section{Domain Wall Approach}
Let us do the analysis of the previous section
(the 4 dim massless Euclidean QED) in the domain wall approach.
First we express the effective action in terms of 
$i\ga_\m(\pl_\m+i\e A_\m)$ itself, not its square as in Sec.2..
\begin{eqnarray}
\Lcal=\psibar\Dvec\psi\com\q
\Dvec=i\ga_\m(\pl_\m+i\e A_\m)\com\q
\Dvec^\dag=\Dvec\pr
\label{DW1}
\end{eqnarray}
Formally we have
\begin{eqnarray}
\ln\,Z[A]=\ln\,
\int\Dcal\psibar\Dcal\psi\e^{-\intfx\Lcal}
=\Tr\ln\Dvec=-\Tr\int_0^\infty\frac{\e^{-t\Dvec}}{t}dt+\mbox{const}\nn
=-\int_0^\infty\frac{dt}{t}\Tr
[\half(1+i\ga_5) \e^{+it\ga_5\Dvec}+
 \half(1-i\ga_5) \e^{-it\ga_5\Dvec}]+\mbox{const}\ .
\label{DW2}
\end{eqnarray}
Because the eigenvalues of $\Dhat$ are both negative and positive,
the $t$-integral above is divergent.
We clearly need regularization to make it meaningful.
We should notice here that the final equality above
relies only on the following properties of $\Dhat$ and $\gago$:
\begin{eqnarray}
\ga_5\Dvec+\Dvec\ga_5=0 \com\q (\gago)^2=1\pr
\label{DW3}
\end{eqnarray}
Note that, in the final expression of (\ref{DW2}),
the signs of the eigenvalues of $\Dhat$ become less
important ( than the case of the previous section )
for the $t$-integral convergence. This is because
the {\it exponetial} operator $\e^{-t\Dhat}$ is replaced
by the {\it oscillating} operators $\e^{\pm it\gago\Dhat}$
due to the relation (\ref{DW3})
\footnote{
This statement is a disguise at the present stage. It is correct
after the {\it Wick rotations} for $t$ at (\ref{DW5}).
Concretely saying, "cos" and "sin" functions will appear in
Subsect.4.2 and Sect. 6(ii).
}
. 
Here we introduce two regularization parameters $M$ and $M'$,
which are most characteristic in this approach.
\footnote{
In a sense we generalize the relation (\ref{DW3})
through this procedure. 
In order to derive eq.(\ref{DW4}) from eq.(\ref{DW2})
for an infinitesimally small value of $M (=M')$, some $M$-dependent terms should
appear in the right hand side of the first equation of (\ref{DW3}).   
It looks to correspond to
a kind of Ginsparg-Wilson relation\cite{GW82}.
}
\begin{eqnarray}
\ln\,Z=-\lim_{M\ra 0}\int_0^{\infty}\frac{dt}{t}
\half(1-i\frac{\pl}{t\,\pl M})\Tr \GfMp(x,y;t)  \nn
-\lim_{M'\ra 0}\int_0^{\infty}\frac{dt}{t}
\half(1-i\frac{\pl}{t\,\pl M'})\Tr \GfMpm(x,y;t)\com\nn
\mbox{where}\nn
\GfMp(x,y;t)\equiv <x|\exp\{+it\ga_5(\Dvec+iM)\}|y>\com\nn
\GfMpm(x,y;t)\equiv <x|\exp\{-it\ga_5(\Dvec+iM')\}|y>
                    \pr
\label{DW4}
\end{eqnarray}
$M$ and $M'$ can be regarded as the "sources" for $\gago$.
Through this procedure we can treat $\gago$ within the new
heat kernels $\GfMp$ and $\GfMpm$.
From its usage above, the limit $M\ra 0, M'\ra 0$ should be taken in the
following way before $t$-integral:
\begin{eqnarray}
|M|t\ll 1\com\q |M'|t\ll 1\pr
\label{DW4b}
\end{eqnarray}
Very interestingly, the above heat-kernels satisfy 
the same (except masses) 1+4 dim
Minkowski Dirac equation after the following {\it Wick rotations} for $t$.
\begin{eqnarray}
(i\dbslash-M)\GfMp=i\e \Aslash \GfMp\com\q
(X^a)=(-it,x^\m)\com                    \nn
(i\dbslash-M')\GfMpm =i\e \Aslash \GfMpm\com\q
(X^a)=(+it,x^\m)                   \com
\label{DW5}
\end{eqnarray}
where $\Aslash\equiv\ga_\m A_\m(x)\ ,\ 
\dbslash\equiv\Ga^a\frac{\pl}{\pl X^a}$. (
$\m$ is the 4 dim Euclidean space indices and runs from 1 to 4, 
while $a$ is the 1+4 dim Minkowski space-time indices and runs 
from 0 to 4. 
A slash '/' is used for 4 dim gamma matrix ($\ga_\m$) contraction, whereas
a backslash '$\backslash$' for 1+4 dim one ($\Ga_a$).  
See App.A for the present notation.) Note here that
the signs of the Wick-rotation is different for $\GfMp$ and $\GfMpm$.
$\GfMp$ and $\GfMpm$ turn out to
correspond to (+)-domain and (-)-domain
in the original formulation \cite{NN94,NN95}
\footnote{
See the explanation in the first paragraph after eq.(\ref{DW13bb}).
}
and we also call them in the same way.

\q Eq.(\ref{DW5}) says $\GfMp$ and $\GfMpm$ are given by the
solutions of the 1+4 dim Dirac equation.
The perturbative solutions are given by the standard textbooks\cite{BD}. 
For simplicity we consider the case $M=M'>0$ in the following.
Both $\GfMp$ and $\GfMm$ are obtained in the same form $G^5_M$
specified by ($G_0,S$).
\begin{eqnarray}
(i\dbslash-M)G^5_{M}=i\e \Aslash G^5_{M}\com\nn
G^5_M(X,Y)=G_0(X,Y)+\int d^5Z\,S(X,Z)i\e\Aslash(z)G^5_M(Z,Y)\com
\label{DW6}
\end{eqnarray}
where $(X^a)=(X^0,X^\m=x^\m)$ and 
$G_0(X,Y)$ is the free solution and $S(X,Z)$ is the propagator:
\begin{eqnarray}
(i\dbslash-M)G_0(X,Y)=0\com\q
(i\dbslash-M)S(X,Y)=\del^5(X-Y)\pr
\label{DW7}
\end{eqnarray}
There are four choices of the above propagator $S(X,Y)$. 
See Fig.1. From them we make three solutions
 and discuss them separately in Subsec.3.1 and 3.2 below.
\begin{figure}
\centerline{\epsfysize=4cm\epsfbox{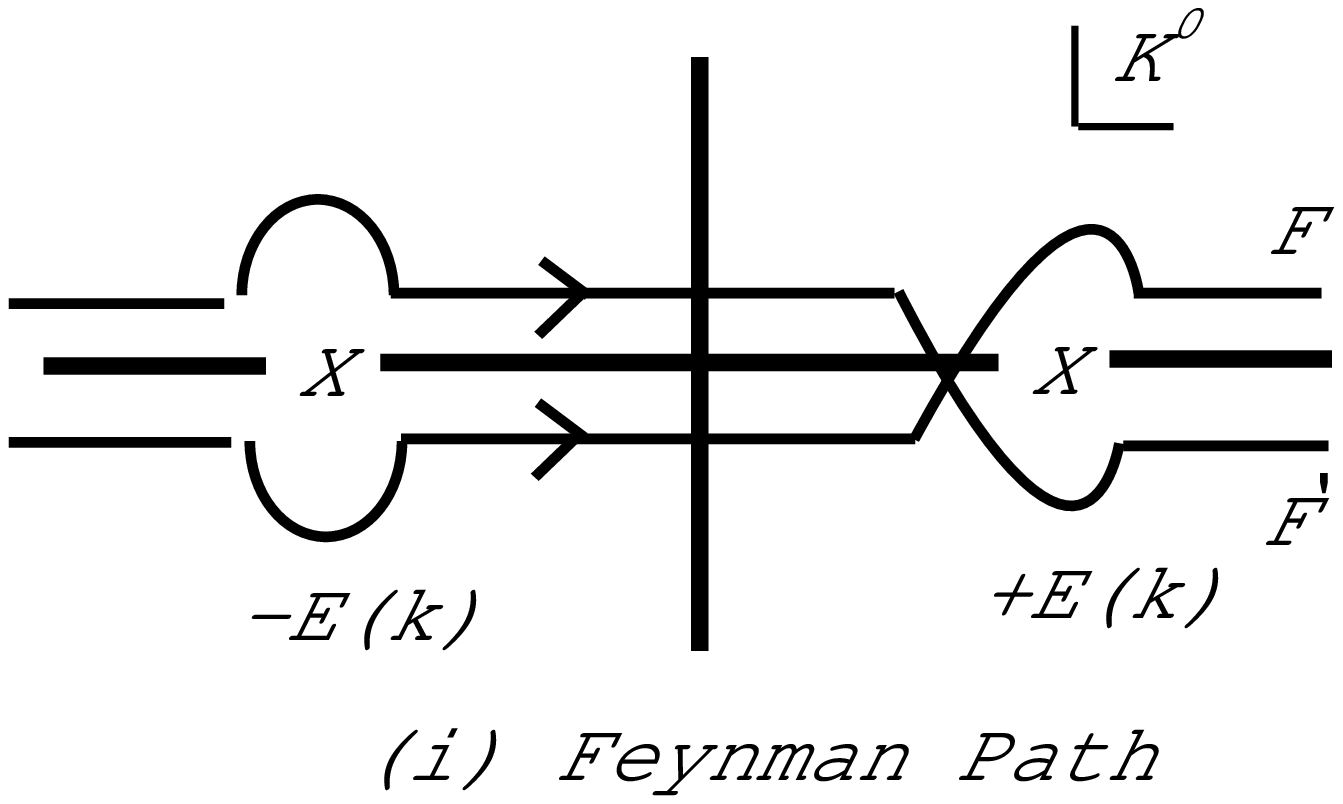}\\
            \epsfysize=4cm\epsfbox{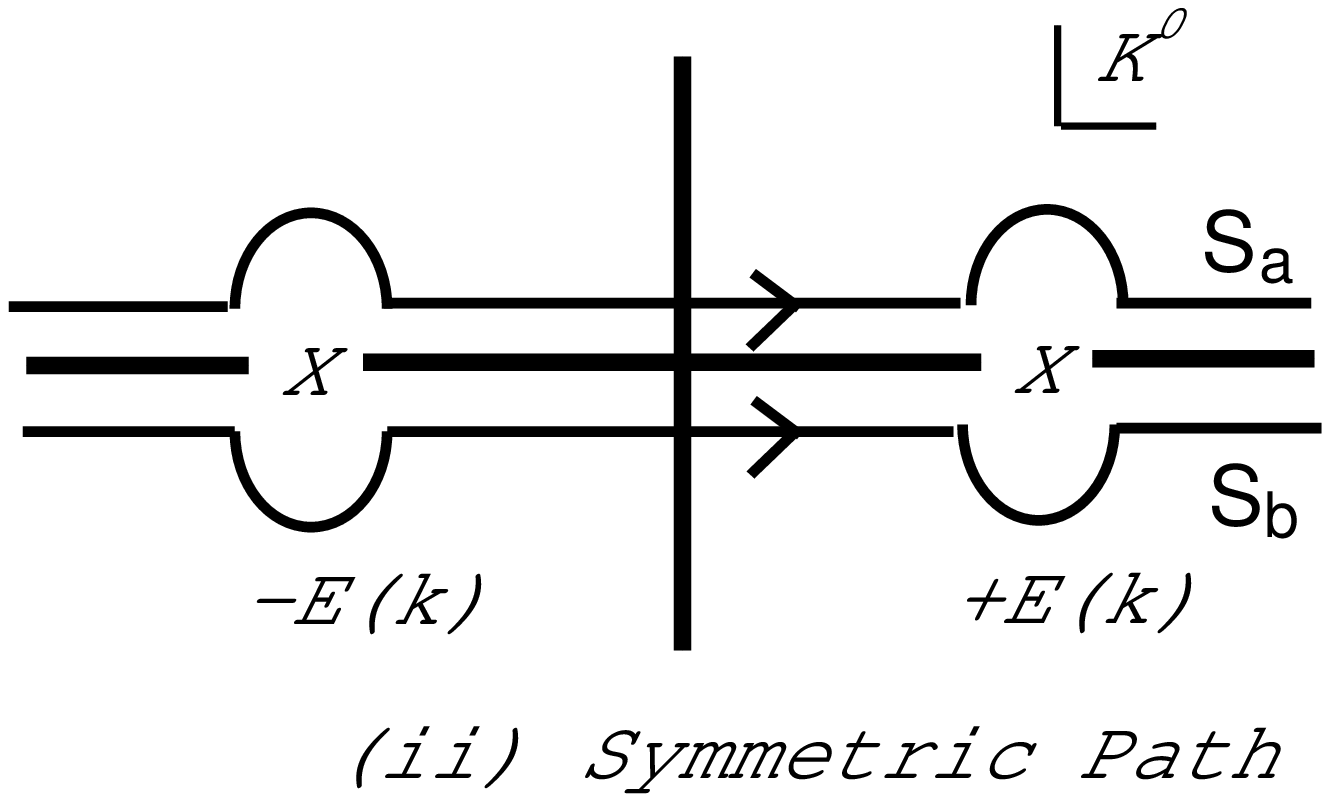}}
   \begin{center}
Fig.1\ 
Four possible pathes for the 1+4 dim Dirac Fermion propagator.
 $E(k)=\sqrt{k^2+M^2}>0.$
   \end{center}
\end{figure}
They are obtained by 
some combinations of the positive and negative
energy free solutions:
\begin{eqnarray}
G^p_0(X,Y)\equiv -i\int\frac{d^4k}{(2\pi)^4}\Om_+(k)\e^{-i\Ktil(X-Y)}
\equiv \int\frac{d^4k}{(2\pi)^4}G^p_0(k)\e^{-ik(x-y)}\ ,\nn 
\Om_+(k)\equiv\frac{M+\Ktilbslash}{2E(k)}\ ,\ 
G^p_0(k)\equiv -i\Om_+(k)\e^{-iE(k)(X^0-Y^0)}              \nn
G^n_0(X,Y)\equiv -i\int\frac{d^4k}{(2\pi)^4}\Om_-(k)\e^{+i\Kbar(X-Y)}
\equiv \int\frac{d^4k}{(2\pi)^4}G^n_0(k)\e^{-ik(x-y)}\ ,\nn  
\Om_-(k)\equiv\frac{M-\Kbarbslash}{2E(k)}\ ,\ 
G^n_0(k)\equiv -i\Om_-(k)\e^{iE(k)(X^0-Y^0)}
\label{DW8}
\end{eqnarray}
where $E(k)=\sqrt{k^2+M^2},(\Ktil^a)=(\Ktil^0=E(k),\Ktil^\m=K^\m=-k^\m),
(\Kbar^a)=$ 
$(\Kbar^0=E(k),\Kbar^\m=-K^\m=k^\m)$. 
$k^\m$ is the momentum in the 4 dim Euclidean space.
\footnote{
The relation between 4 dim quantities($x^\m$ and $k^\m$)
and 1+4 dim ones($X^a$ and $K^a$):\ 
$(X^a)=(X^0,X^\m=x^\m), (K^a)=(K^0,K^\m=-k^\m),
K_aX^a=K_0X^0-K^\m X^\m=K_0X^0+k^\m x^\m.$
Only for 1+4 dim quantities (capital letters), the upper
and lower indices have meaning.
}
($\m=1,2,3,4;\ a=0,1,2,3,4$.)\ 
$\Ktil$ and $\Kbar$ are on-shell
momenta($\Ktil^2=\Kbar^2=M^2$), which correspond to the positive
and negative energy states respectively.
\footnote{
Useful relations:\ 
$
-i\Ktil X=-iE(k)X^0-ikx,\ i\Kbar X=iE(k)X^0-ikx,\ 
M+\Ktilbslash=M+E(k)\gago+i\kslash,\ 
M-\Kbarbslash=M-E(k)\gago+i\kslash.
$
}
We can rewrite $\Om_\pm(k)$ as
\begin{eqnarray}
\Om_+(k)=\half (\gago+\frac{M+i\kslash}{|M+i\kslash|})\com\q 
\Om_-(k)=\half (-\gago+\frac{M+i\kslash}{|M+i\kslash|}) 
\label{DW8.5}
\end{eqnarray}
where $|M+i\kslash|\equiv E(k)=\sqrt{k^2+M^2}$. The expressions above
look similar to the overlap Dirac operator
\cite{Neub98} for the case of {\it no} Wilson term.
\footnote{
The overlap Dirac operator $D$ on lattice is\nl 
$ S_F=a^4\sum_{x\in \mbox{all sites}}\psibar(x)D\psi(x),\ 
aD=1+\gago\frac{H}{\sqrt{H^2}},\ $\nl
$\gago H=\sum_\m \{ \half \ga_\m(\na_\m+\na_\m^*)
-\frac{a}{2}\na_\m^*\na_\m \}-M,$\nl
where $a$ is a lattice spacing. If we ignore the 
$\na_\m^*\na_\m$-term (Wilson term), the form is
quite similar to (\ref{DW8.5}).
}
In fact, $\Om_\pm(k)$ have {\it projective property} with their hermite
conjugate. The relations are listed in App.C and will
be efficiently used in the calculations in Sec.4 and 6.

\q The final important stage is regularization of the (1-loop) 
ultraviolet divergences. (We will explain it in Sec.5 in detail.)
Corresponding to the 1-loop quantum evaluation, the determinant (\ref{DW2})
finally involves one momentum($k^\m$)-integral (besides $t$-integral).
We will take the analytic continuation method (Sect.5(ii)) in order to avoid
introducing further regularization parameters and to avoid breaking the
gauge invariance. As will be explained in (\ref{MI.2b}) in Sect.5(i),  
it is essentially equivalent
to restricting the integral region from $0\leq |k^\m| < \infty$ to
\begin{eqnarray}
\mbox{Chiral Condition}\ :\ 
0\leq |k^\m| \leq M\pr
\label{DW8a}
\end{eqnarray}
(See Sec.5.)
\footnote{
Instead of the analytic continuation, we can take the higher derivative
regularization. This corresponds to the Wilson term in lattice:\ 
$i\dslash -M\ \ra\ i\dslash\pm\frac{r}{M}\pl^2-M,$ 
$r$: "Wilson term" coefficient. In this case the unitarity problem,
rather than the gauge invariance, should be clarified. 
}
This looks similar to the usual Pauli-Villars procedure in the point of
ultra-violet regularization. $M$ plays the role of the momentum cut-off.
We should stress that this restriction condition (\ref{DW8a})
on the momentum integral, at the same time, controls the chirality
as explained in the following. ( This point is a distinguished property
of the domain wall regularization. ) We call (\ref{DW8a}) {\it chiral condition}.
In fact, taking the {\it extreme} chiral limit:
\begin{eqnarray}
\mbox{Extreme Chiral Limit}\ :\ 
\frac{M}{|k^\m|}\ra \infty\com
\label{DW8aa}
\end{eqnarray}
in the present case implies the chirality selection.
In the original temperature coordinate $t$ ($X^0-Y^0=\mp it$), 
$G^p_0(k)$ and $G^n_0(k)$ behave as, in the extreme
chiral limit (e.c.l.)\ $M/|k^\m|\ra +\infty$,
\begin{eqnarray}
\mbox{for}\ \mbox{(+)-domain}\q\q (X^0=-it)\qqq\qqq\nn
iG^p_0(k)\ra  \frac{1+\gago}{2}\e^{-Mt}\ \ , \ \ 
iG^n_0(k)\ra  \frac{1-\gago}{2}\e^{+Mt}\ ,\nn
\mbox{for}\ \mbox{(-)-domain}\q\q (X^0=+it)\qqq\qqq\nn
iG^p_0(k)\ra  \frac{1+\gago}{2}\e^{+Mt}\ \ , \ \ 
iG^n_0(k)\ra  \frac{1-\gago}{2}\e^{-Mt}\ \ . 
\label{DW8b}
\end{eqnarray}
This result will be used for characterizing different configurations
with respect to the chirality.
We use (\ref{DW8a}) instead of (\ref{DW8aa}) in concrete
calculations. (\ref{DW8aa}) is {\it too restrictive} to keep the dynamics. 
Loosening the extreme chiral limit (\ref{DW8aa}) to the chiral condition
(\ref{DW8a}) can be regarded as a part of the present regularization.
This situation looks similar to the introduction of the Wilson term, 
in the lattice formalism, in order
to break the chiral symmetry. (See the last paragraph of Sec.8.)

Let us reexamine the condition (\ref{DW4b}). As read from
the above result, the domain is characterized by the exponetial
damping behaviour which
has the "width"$\sim 1/M$ around the
origin of the extra $t$-axis. (\ref{DW4b}) restricts the region
of $t$ as $t\ll 1/M$. This is for considering only the massless mode
as purely as possible.
In the lattice formalism, this corresponds to taking
the zero mode (surface state) limit, in order to avoid the doubling problem,
by introducing many "flavor" fermions (or adding an extra dimension)
and many bosonic Pauli-Villars fields to kill the heavy fermions contribution. 
Besides the extreme chiral limit (\ref{DW8aa}), we often consider 
, corresponding to (\ref{DW4b}), the following limit:\  
\begin{eqnarray}
M|X^0-Y^0|\ra +0\pr
\label{DW8bx}
\end{eqnarray}
$G^p_0(X,Y)$ and $G^n_0(X,Y)$ behave, in this limit, as
\begin{eqnarray}
iG^p_0(X,Y)\ra  \int\frac{d^4k}{(2\pi)^4}\Om_+(k)\e^{-ik(x-y)}\ ,\ 
iG^n_0(X,Y)\ra  \int\frac{d^4k}{(2\pi)^4}\Om_-(k)\e^{-ik(x-y)}\ ,\nn
i(G^p_0(X,Y)-G^n_0(X,Y))\ra  \gago\del^4(x-y)\ ,\nn
i(G^p_0(X,Y)+G^n_0(X,Y))\ra  \int\frac{d^4k}{(2\pi)^4}\frac{M+i\kslash}{|M+i\kslash|}\e^{-ik(x-y)}\ .
\label{DW8c}
\end{eqnarray}
The factor 
$\frac{M+i\kslash}{|M+i\kslash|}$ can be regarded as a "phase" operator 
{\it depending on configuration}.
The above result will be used to find the {\it boundary conditions} of the
full solutions (\ref{DW6}).

\q In the lattice numerical simulation, 
the best fit value of the regularization mass $M$ looks
restricted both from the below and from the above
depending on the simulation "environment"
\cite{BS97PR,Col99}.
\footnote{
In lattice 
the corresponding bound on $M$, from the requirement of no doublers,
has been known since the original works\cite{Kap92,NN94}.
}
($M\sim$a few Gev for the hadron simulation.)
The similar one occurs in the present regularization.
The "double" limits (\ref{DW4b}) and 
(\ref{DW8a}) or (\ref{DW8aa}) imply
\begin{eqnarray}
|k^\m|\ll M\ll \frac{1}{t}\q \mbox{or}\q
|k^\m|\leq M\ll \frac{1}{t} \pr
\label{DW14}
\end{eqnarray}
In the standpoint of the extra dimension, the limit $M\ll \frac{1}{t}$ 
($Mt\ra +0$) corresponds to
, combined with the condition on $|k^\m|/M$, 
taking the dimensional reduction from
1+4 dim to 4 dim (Domain wall picture of 4 dim space).
In the regularization view, this limit plays the role of
the infrared regularization. 
The condition $|k^\m|\leq M$ should be basically regarded as
a control of chirality as explained above.
Its role, from the view of regularization, is 
the control of the high momentum region
in the divergent momentum integral. 

The relation (\ref{DW14}) is the most characteristic one of the 
present regularization. It should be compared with the usual heat-kernel
regularization in (\ref{QED9}) and (\ref{QED10}) where only
the limit $t\ra +0$ is taken and the ultraviolet regularization
is done by the simple subtraction of divergences. 
Eq.(\ref{DW14}) shows the delicacy in taking the limit in the
the present 1+4 dimensional
regularization scheme. 
It implies, in the lattice simulation, 
$M$ should be appropriately chosen depending
on the regularization scale (,say, lattice size) and the  
momentum-region of 4 dim fermions.

\vs 1
\subsection{Feynman Path}  
In this subsection and the next, we obtain some solutions of
1+4 dim Dirac equation which are specified by $(G_0,S)'$s through
the general form (\ref{DW6})-(\ref{DW7}). 
First we consider the Feynman path (F) in Fig.1. Then the propagator
is given by the Feynman propagator:\ 
$S_F(X,Y)=\th (X^0-Y^0)G^p_0(X,Y)+\th(Y^0-X^0)G^n_0(X,Y)$.
It has both the retarded and advanced parts.
Now we remind ourselves of the fact that there exists
a {\it fixed direction} in the system evolvement when the temperature
parameter works well (See the statement around (\ref{QED4b}) 
and (\ref{QED4bb}) in Sec.2).
Let us regard the extra axis, after the Wick-rotations,
as a temperature.  
Assuming the analogy holds here, we try to
adopt the following solution, imitating the form of the Weyl
anomaly solution (\ref{QED4b}).
\begin{eqnarray}
\mbox{Retarded solution for}\q \GfMp:\q \nn
G_0(X,Y)=G^p_0(X,Y)\com\q S(X,Y)=\th (X^0-Y^0)G^p_0(X,Y)
\equiv S_F^+(X,Y)\ ;\label{DW9a}\\
\mbox{Advanced solution for}\q \GfMm:\q \nn
G_0(X,Y)=G^n_0(X,Y)\com\q S(X,Y)=\th (Y^0-X^0)G^n_0(X,Y)
\equiv S_F^-(X,Y)\ \label{DW9b}
\end{eqnarray}
This is chosen in such a way that the $t$-integral converges.
(The "opposite" choice will be considered in the last part of this subsection.)
Because we have "divided" a full solution into two chiral parts in order to introduce
a {\it fixed direction} in the system evolvement, $S_F^\pm$ above
does not satisfy the proper propagation equation (\ref{DW7}).
Instead it satisfies the following ones:
\begin{eqnarray}
(i\dbslash-M)S^+_F=\frac{1+\gago}{2}\del^5(X-Y)\nn
+\gago\del (X^0-Y^0)
\int\frac{d^4k}{(2\pi)^4}(\frac{i\kslash}{2M}+O((\frac{k}{M})^2) )
\e^{-ik(x-y)}\com\nn
(i\dbslash-M)S^-_F=\frac{1-\gago}{2}\del^5(X-Y)\nn
-\gago\del (X^0-Y^0)
\int\frac{d^4k}{(2\pi)^4}(\frac{i\kslash}{2M}+O((\frac{k}{M})^2) )
\e^{-ik(x-y)}\com
\label{DW9bx}
\end{eqnarray}
The equations above say $S_F^\pm$ above satisfy
the "chiral" propagator equation 
at the extreme chiral limit:\ 
$M/|k^\m|\ra +\infty$.
$G^{5M}_\pm$ defined by (\ref{DW9a},\ref{DW9b}), through the second equation of (\ref{DW6}), 
do not satisfy 
$(i\dbslash-M)G^5_{M}=i\e \Aslash G^5_{M}$ but satisfy
\begin{eqnarray}
M\ra +\infty\com\nn
(i\dbslash-M)\GfMp =i\e \frac{1+\gago}{2}\Aslash \GfMp +O(\frac{1}{M})\com\nn
(i\dbslash-M)\GfMm =i\e \frac{1-\gago}{2}\Aslash \GfMm +O(\frac{1}{M})
\pr \label{DW9by}
\end{eqnarray}
They correspond to the determinant of the {\it chiral} QED:\ 
$\Dhat_\pm\equiv i(\dslash+ie\frac{1\pm\gago}{2}\Aslash)$\ 
instead of $\Dhat$ of (\ref{DW1}). 
Taking the extreme chiral limit in the momentum spectrum of
the propagators $S_F^\pm$ (\ref{DW9a},\ref{DW9b}), we can read off
the domain wall structure as in Fig.2.
\begin{figure}
\centerline{\epsfysize=6cm\epsfbox{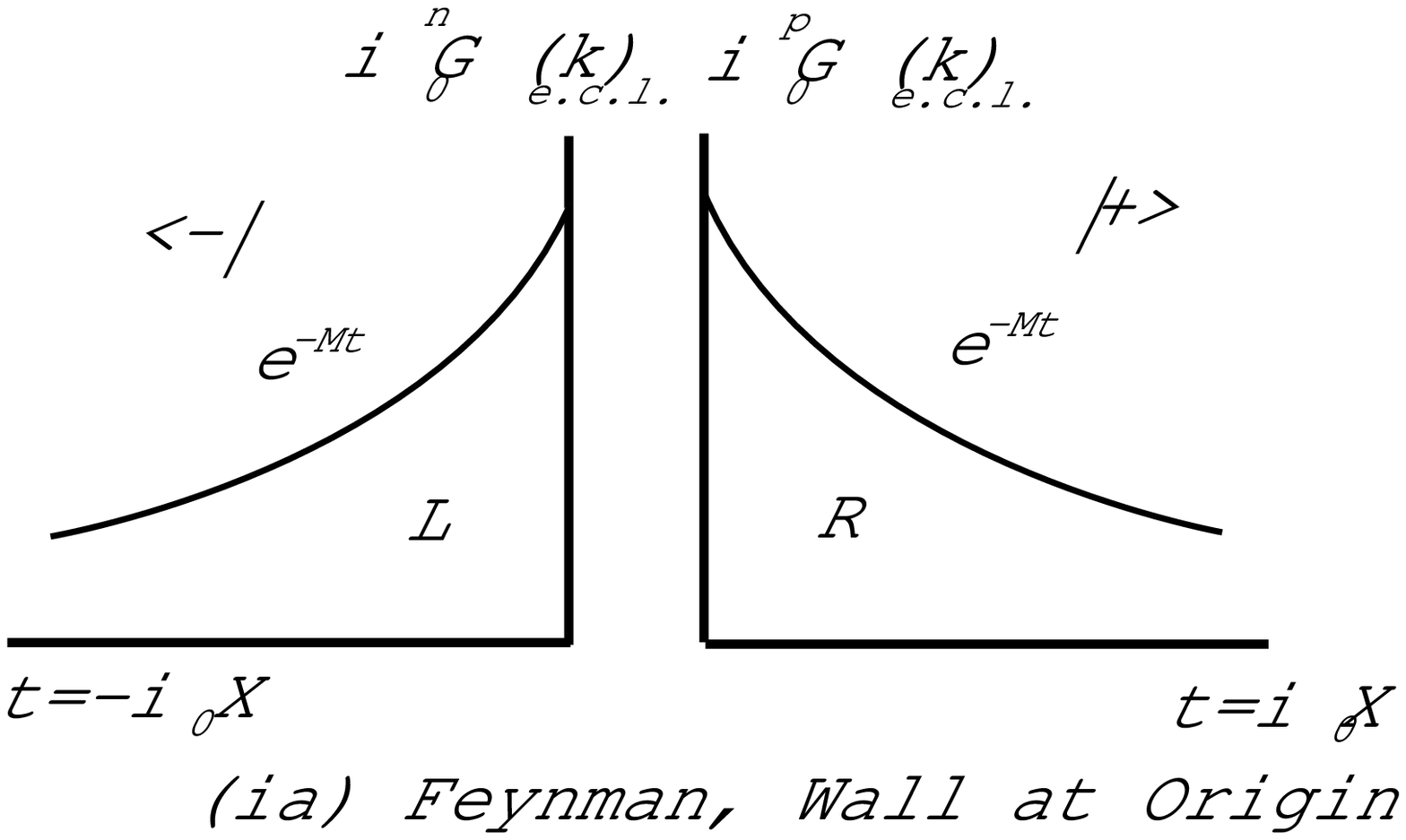}}
   \begin{center}
Fig.2\ 
Domain Wall structure read from the momentum spectrum of
the Feynman path propagators (\ref{DW9a},\ref{DW9b}) at the
extreme chiral limit (e.c.l.) (\ref{DW8aa}). 
$S_F^+$ for $|+>$ and $S_F^-$ for $|->$. 
   \end{center}
\end{figure}
The full solutions ${G^{5M}_\pm}$ (\ref{DW6}), made by
the solutions (\ref{DW9a},\ref{DW9b}), satisfy 
the same boundary condition as the free ones (\ref{DW8c})
from its construction:
\begin{eqnarray}
\GfMp(X,Y)\ \mbox{(Retarded)}\ra -i\int\frac{d^4k}{(2\pi)^4}\Om_+(k)\e^{-ik(x-y)}
\ \mbox{as}\ M(X^0-Y^0)\ra +0\ ,\nn
\GfMm(X,Y)\ \mbox{(Advanced)}\ra -i\int\frac{d^4k}{(2\pi)^4}\Om_-(k)\e^{-ik(x-y)}
\ \mbox{as}\ M(X^0-Y^0)\ra -0\ .
\label{DW9c}
\end{eqnarray}
From this we obtain
\begin{eqnarray}
i(\GfMp(X,Y)-\GfMm(X,Y))\ra \gago\del^4(x-y)\ \mbox{as}\ 
M|X^0-Y^0|\ra +0\ ,\nn
i(\GfMp(X,Y)+\GfMm(X,Y))\ra 
\int\frac{d^4k}{(2\pi)^4}\frac{M+i\kslash}{|M+i\kslash|}\e^{-ik(x-y)}
\ \mbox{as}\ M|X^0-Y^0|\ra +0\ ,
\label{DW9d}
\end{eqnarray}
where $|M+i\kslash|\equiv E(k)$. 
Taking into account
the boundary conditions above, we should take
, in the anomaly calculation (\ref{QED9},\ref{QED10}), as
\begin{eqnarray}
\half\del_\al\,\ln\,J_{ABJ}=\lim_{M|X^0-Y^0|\ra +0}
\Tr\,i\al(x)i(\GfMp(X,Y)-\GfMm(X,Y))\com\nn
\half\del_\om\,\ln\,J_W=\lim_{M|X^0-Y^0|\ra +0}
\Tr\,\om(x)i\gago (\GfMp(X,Y)-\GfMm(X,Y))\pr
\label{DW10}
\end{eqnarray}

\q The meaning of the choice of Feynman path solution
(\ref{DW9a},\ref{DW9b})
is subtle (but interesting)
\footnote{
As for the possibility of defining the chiral theory using
this solution, see the argument in the second paragraph of Sect.8.
}
, because the solution does not satisfy 
$(i\dbslash-M)G^5_{M}=i\e \Aslash G^5_{M}$. The clear separation
of right and left (Fig.2) and its calculational simplicity fascinate
us to examine this solution. One purpose of this paper is to find whether
this solution works correctly as regularization or not. ( As will be seen
in the following sections, it works well except a simple factor
as far as anomalies are concerned. ) The same thing can be said
about the next paragraph.

\q As a final comment of this subsection, we refer to the opposite choice
in (\ref{DW9a}) and (\ref{DW9b}).
We call this choice anti-Feynman path solution. 
If we take the anti-Feynman (F') path in Fig.1, we obtain
\begin{eqnarray}
\mbox{Retarded solution for}\q \GfMp:\q \nn
G_0(X,Y)=G^n_0(X,Y)\com\q S(X,Y)=\th (X^0-Y^0)G^n_0(X,Y) \label{DW12c} \\
\mbox{Advanced solution for}\q \GfMm:\q \nn
G_0(X,Y)=G^p_0(X,Y)\com\q S(X,Y)=\th (Y^0-X^0)G^p_0(X,Y)
\label{DW12d}
\end{eqnarray}
Taking the extreme chiral limit $|\frac{k^\m}{M}|\ll 1$  
in the chosen propagator above, we can read
the domain wall structure as in Fig.3.
\begin{figure}
\centerline{\epsfysize=6cm\epsfbox{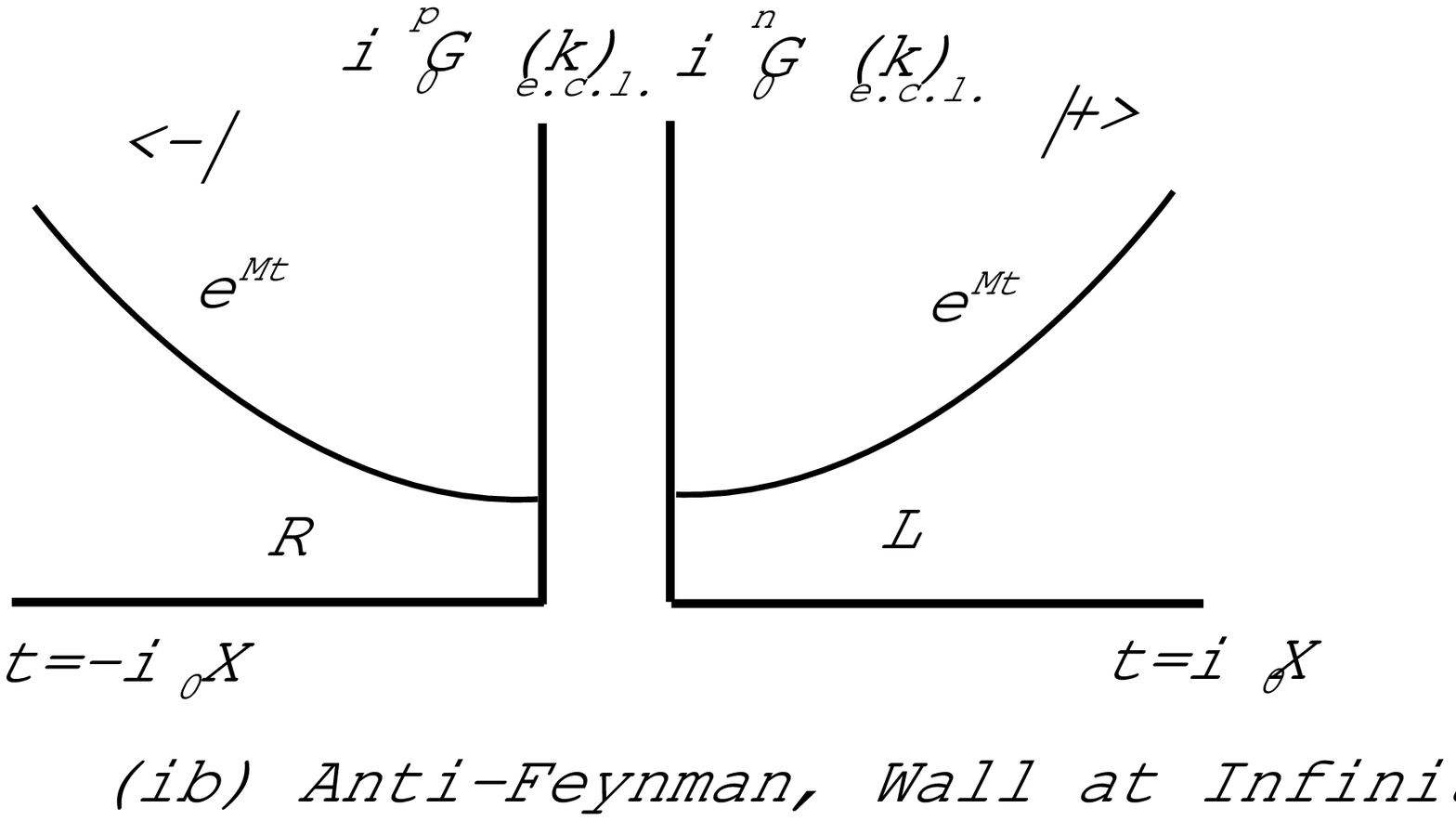}}
   \begin{center}
Fig.3\ 
Domain Wall structure read from the momentum spectrum of
the anti-Feynman path propagators 
(\ref{DW12c},\ref{DW12d}) at the extreme chiral limit (e.c.l.) (\ref{DW8aa}).
   \end{center}
\end{figure}
It shows the domain wall not at the origin ($t=0$) but at
the infinity ($t=\infty$) for each domain.
The above solutions satisfy the boundary condition:
\begin{eqnarray}
\GfMp(X,Y)\ \mbox{(Retarded)}\ra -i\int\frac{d^4k}{(2\pi)^4}\Om_-(k)\e^{-ik(x-y)}
\ \mbox{as}\ M(X^0-Y^0)\ra +0\ ,\nn
\GfMm(X,Y)\ \mbox{(Advanced)}\ra -i\int\frac{d^4k}{(2\pi)^4}\Om_+(k)\e^{-ik(x-y)}
\ \mbox{as}\ M(X^0-Y^0)\ra -0\ .
\label{DW12e}
\end{eqnarray}
From this we obtain
\begin{eqnarray}
i(\GfMp(X,Y)-\GfMm(X,Y))\ra -\gago\del^4(x-y)\ \mbox{as}\ 
M|X^0-Y^0|\ra +0\ ,\nn
i(\GfMp(X,Y)+\GfMm(X,Y))\ra 
\int\frac{d^4k}{(2\pi)^4}\frac{M+i\kslash}{|M+i\kslash|}\e^{-ik(x-y)}
\ \mbox{as}\ M|X^0-Y^0|\ra +0\ .
\label{DW12f}
\end{eqnarray}
The regularization using this solution turns out to
give the same result as the Feynman path solution. 
The different point is 
that, due to the presence of the exponetially growing
factor $\e^{+E(k)t}$, we must do calculation in
the $X^0$-coordinate.\cite{SI98}

\subsection{Symmetric Path}

Let us consider the symmetric pathes $S_a$ and $S_b$ in Fig.1.
In this case we are led to take the following solution.
\begin{eqnarray}
\mbox{Symmetric retarded solution for}\q \GfMp:\q \nn
G_0(X,Y)=G^{\mbox{p}}_0(X,Y)-G^{\mbox{n}}_0(X,Y)\ ,\nn 
S(X,Y)=\th (X^0-Y^0)(G^{\mbox{p}}_0(X,Y)-G^{\mbox{n}}_0(X,Y))
\equiv S^+_{sym}(X,Y);     \label{DW13a}\\
\mbox{Symmetric advanced solution for}\q \GfMm:\q \nn
G_0(X,Y)=G^{\mbox{n}}_0(X,Y)-G^{\mbox{p}}_0(X,Y)\ ,\nn 
S(X,Y)=\th (Y^0-X^0)(G^{\mbox{n}}_0(X,Y)-G^{\mbox{p}}_0(X,Y))
\equiv S^-_{sym}(X,Y)\pr\label{DW13b}
\end{eqnarray}
$S^\pm_{sym}$ satify the propagator equation properly,
\begin{eqnarray}
(i\dbslash-M)S^\pm_{sym}=\del^5(X-Y)\com
\label{DW13bw}
\end{eqnarray}
which should be compared with $S^\pm_F$ of (\ref{DW9bx}).
\footnote{
The proper solution of (\ref{DW6}) with the initial condition (\ref{DW13bx}),
which is the Cauchy problem of 1+4 dim Dirac equation, is given here.
Note, in the present case, that 
the time axis is a half line $X^0>0$ or $X^0<0$ with $Y^0=0$.
With this note, the results (\ref{DW13a}) and (\ref{DW13b})
coincide with 1+4 dim version of (3.29) of \cite{AN92}.
}
Taking the extreme chiral limit $\frac{|k^\m|}{M}\ll 1$  
in the spectrum of the chosen propagator above, we can read off
the symmetric wall structure 
(one wall at the origin and the other at the infinity) as in Fig.4.
\begin{figure}
\centerline{\epsfysize=6cm\epsfbox{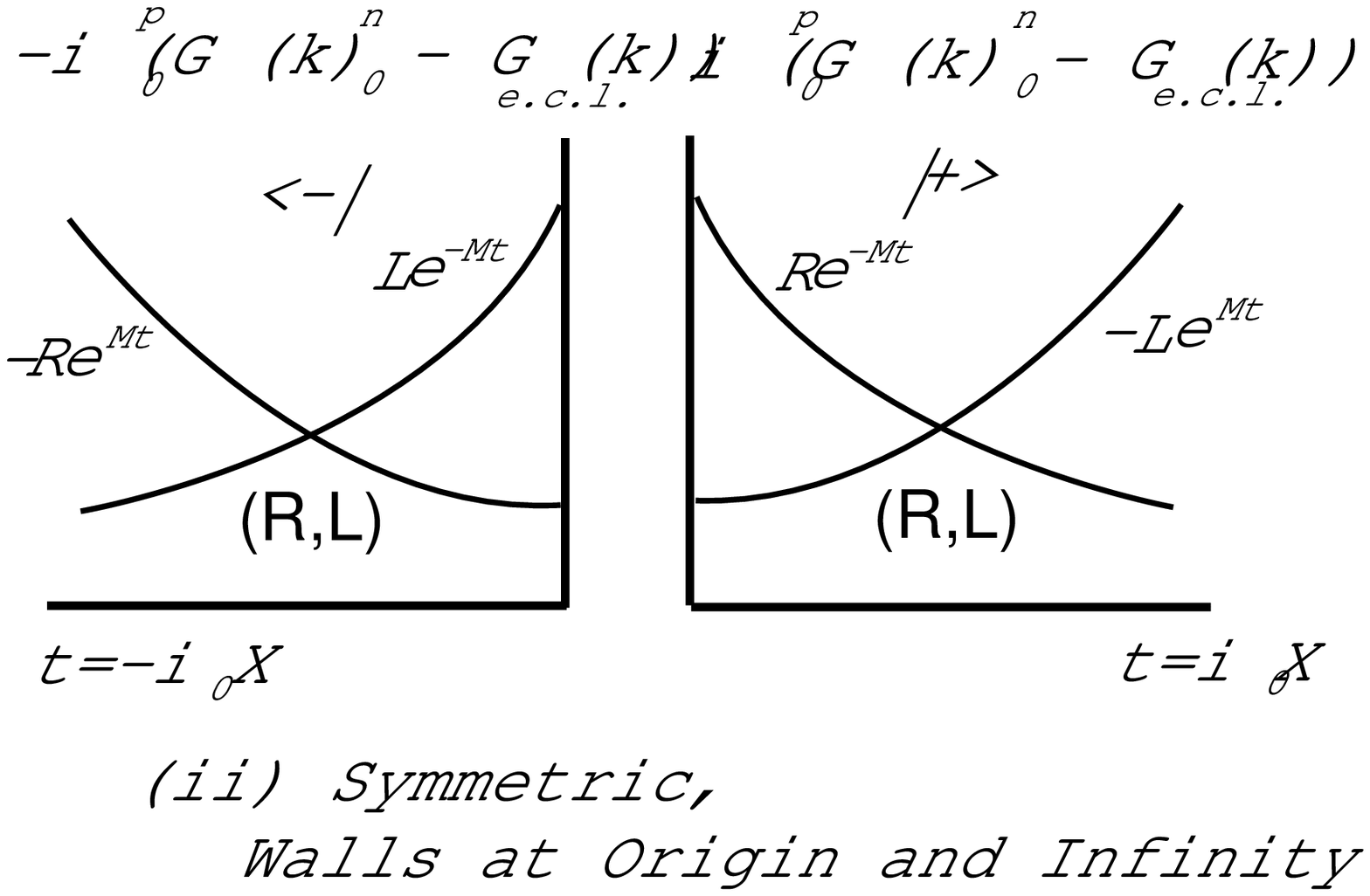}}
   \begin{center}
Fig.4\ 
Domain Wall structure read from 
the momentum spectrum of the Symmetric path propagators 
(\ref{DW13a},\ref{DW13b}) at the extreme chiral limit (e.c.l.) (\ref{DW8aa}).
$S^+_{sym}$ for $|+>$ and $S^-_{sym}$ for $|->$
   \end{center}
\end{figure}
The above solutions satisfy the following boundary condition:
\begin{eqnarray}
\GfMp\mbox{(Retarded)}\ra -i\gago\del^4(x-y)\ \mbox{as}\ 
M(X^0-Y^0)\ra +0\ ,\nn
\GfMm\mbox{(Advanced)}\ra +i\gago\del^4(x-y)\ \mbox{as}\ 
M(X^0-Y^0)\ra -0\ ,\nn
\frac{i}{2}(\GfMp-\GfMm)\ra \gago\del^4(x-y)\ \mbox{as}\ 
M|X^0-Y^0|\ra +0\ .
\label{DW13bx}
\end{eqnarray}

In this case, the measure change (\ref{QED9},\ref{QED10}) 
is regularized as 
\begin{eqnarray}
\del_\al\,\ln\,J_{ABJ}=
\lim_{M(X^0-Y^0)\ra +0}\Tr\,i\al(x)i\GfMp(X,Y)\nn
+\lim_{M(X^0-Y^0)\ra -0}\Tr\,i\al(x)(-i)\GfMm(X,Y)\nn
=\lim_{M|X^0-Y^0|\ra +0}\Tr\,i^2\al(x)\{\GfMp(X,Y)-\GfMm(X,Y)\}
\com\nn
\del_\om\,\ln\,J_{W}=
\lim_{M(X^0-Y^0)\ra +0}\Tr\,\om(x)i\gago\GfMp(X,Y)\nn
+\lim_{M(X^0-Y^0)\ra -0}\Tr\,\om(x)(-i)\gago\GfMm(X,Y)\nn
=\lim_{M|X^0-Y^0|\ra +0}\Tr\,i\om(x)\gago\{\GfMp(X,Y)-\GfMm(X,Y)\}
\pr
\label{DW13bb}
\end{eqnarray}

\vspace{1cm}
The figures of Fig.2,3 and 4 schematically show the present regularization
can control the chirality well because the separation between
the left and the right is the key point. 
Both in (\ref{DW10}) and in (\ref{DW13bb}),
the anomalies are expressed by the "difference" between $\GfMp$ 
and $\GfMm$ contributions. 
\footnote{
Especially for the Feynman path case (\ref{DW10}),
taking the difference is indispensable to regularize
$\gago \del^4(x-y)$.
}
This exactly corresponds to the "overlap" equation in the
original formalism. The "difference" in the effective action corresponds
to the "product" in the partition function between (+) part and (-) part,
that is the "overlap". This is the reason we can name $G^{5M}_\pm$
as ($\pm$)-domains in (\ref{DW4}) and (\ref{DW5}).

\q So far we have mainly explained the formalism of the domain wall
regularization.
In the following sections, we will explicitly evaluate the chiral anomalies
in 4 dim QED and 2 dim chiral gauge theory
using this new domain wall formalism. Surely the known
results are reproduced. This shows the correctness of the present
regularization. Some different regularizations appear depending
on the choice of solutions of the 1+4 dim massive Dirac equation.
Each choice has its characteristic aspect. The Feynman and anti-Feynman
pathes are advantageous in that the calculation is simple. Especially for
the Feynman path, 
the evaluation can be done in the original $t$-coordinate
(no need for Wick rotation).
The clear separation of left and right chirality
is also advantageous. 
The chiral version of the original theory
, which is non-chiral (hermitian), is automatically treated
at the limit $M\ra +\infty$. 
It is disadvantageous(, at least, at present )
that the regularization relies
on the "approximate" solution of the Dirac equation 
as shown in (\ref{DW9bx}). 
On the other hand
the symmetric path is advantageous in that it relies on
the proper solution of the Dirac equation as shown in (\ref{DW13bw}). 
The configuration
of two walls (one at the origin and the other at the infinity 
for each domain)
\footnote{
In lattice, the wall at the infinity is often called "anti-wall".
This is the chiral partner of the other at the origin in the
symmetric solution. Do not confuse it with the similar terminology
:\ the "anti-Feynman" path in Subsec.3.1.
}
is similar to the lattice situation. 
Its disadvantageous point is
the calculational complexity. 
We must take into account both positive and negative energy states
for every propagator.
We will later see, in Sec.6, another
important difference, between the above two kinds pathes, 
in relation to the consistent and covariant anomalies.

\section{Anomalies of 4 Dim Euclidean QED Using Domain Wall Regularization}
In this section we explicitly evaluate chiral 
anomalies.
In the process some divergent integrals will appear. They correspond to
the ultraviolet divergences in the local field theories.
Its regularization is one of key points of the present approach
and is separately examined in Sec.5.
Only the 2-nd order (with respect to $A_\m$) perturbation contributes.

\subsection{Feynman Path}
The first term of (\ref{DW10}) is evaluated as,
taking $X_0>Y_0\equiv 0$, 
\begin{eqnarray}
\GfMp|_{AA}
=\int^{X^0}_0dZ^0\int^{Z^0}_0dW^0
\int d^4Z\int d^4W   \nn
G^p_0(X,Z)i\e \Aslash(z)G^p_0(Z,W)
i\e \Aslash(w)G^p_0(W,Y)\nn
=\int^{X^0}_0dZ^0\int^{Z^0}_0dW^0
\int d^4z\int d^4w\int\frac{d^4k}{(2\pi)^4}\frac{d^4l}{(2\pi)^4}
\frac{d^4q}{(2\pi)^4}\nn
\times (-i)\Om_+(k)i e \Aslash(z)
(-i)\Om_+(l) i e \Aslash(w)(-i)\Om_+(q)\nn
\times\exp \{-i\Ktil(X-Z)-i\Ltil(Z-W)-i\Qtil(W-Y)\}
\com
\label{DW11}
\end{eqnarray}
where $\Om_+(k)\equiv\frac{M+\Ktilbslash}{2E(k)}$.
See Fig.5(i).
\begin{figure}
\centerline{\epsfysize=6cm\epsfbox{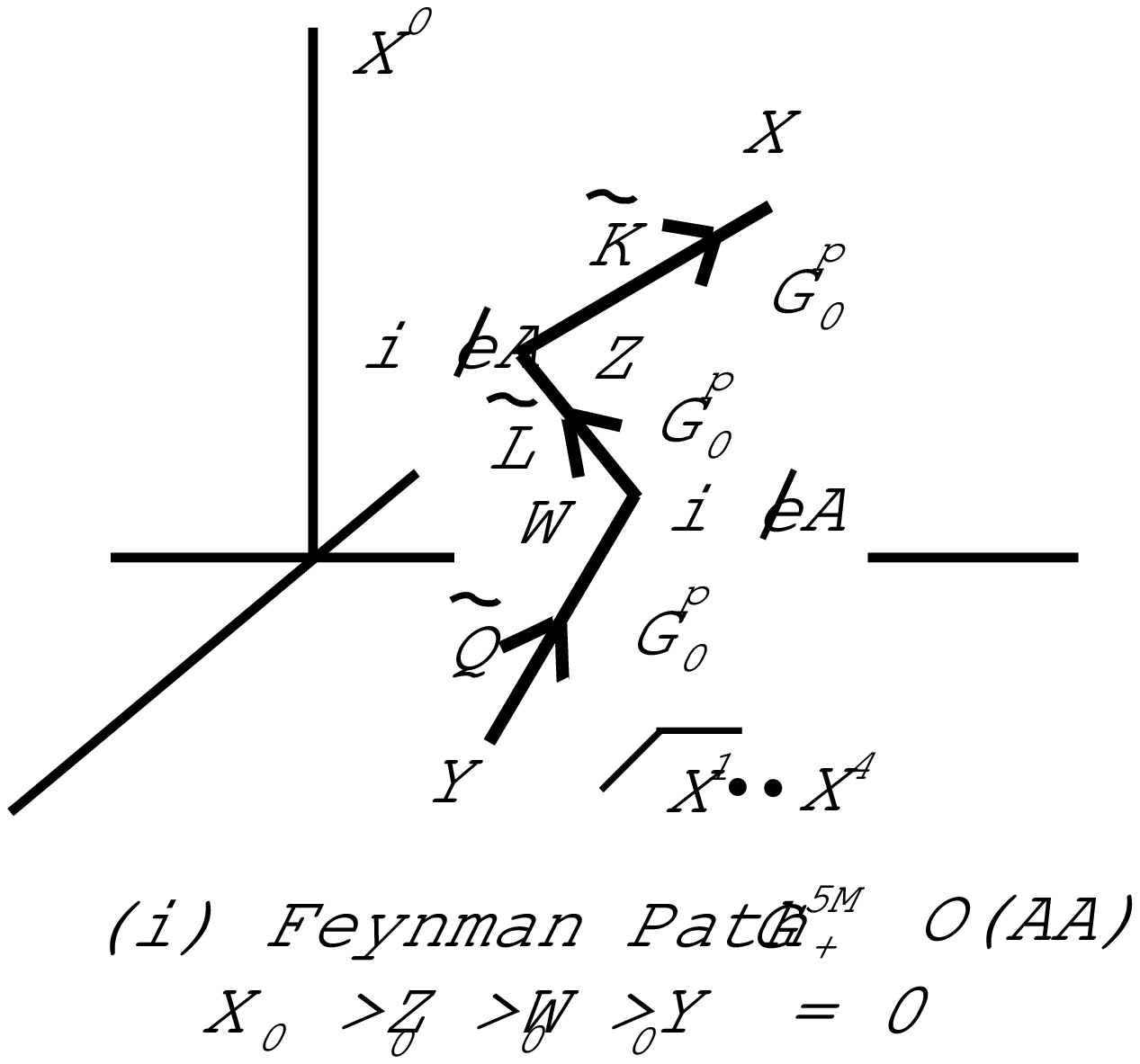}\\
            \epsfysize=6cm\epsfbox{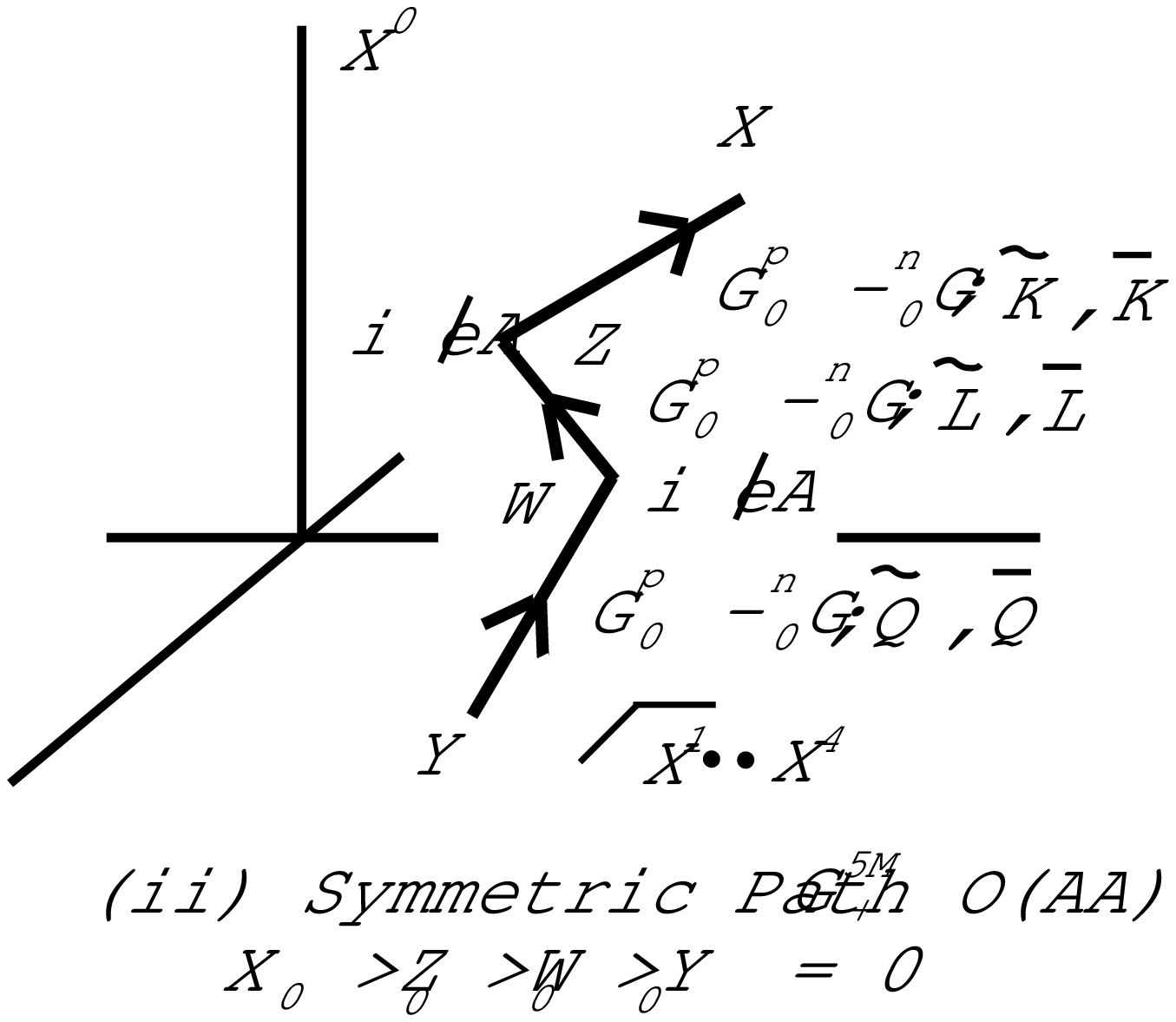}}
   \begin{center}
Fig.5\ 
Abelian gauge theory, $\GfMp$, $O(AA)$, (i) Feynman path and (ii) symmetric path.
   \end{center}
\end{figure}
As we see from the lower-ends of $Z^0$ and $W^0$-integrals, 
we have made here an important assumption about the extra axis:\ 
the axis is a {\it half line} ( not a (straight) line ) like the temperature ($t$) 
axis of Sec.2.
\footnote{
If we take the extra axis as a (straight) line, 
we have to introduce an {\it infrared}
cut-off in the $Z^0$ and $W^0$-integrals in (\ref{DW11}), 
and the final anomaly result depends on it.
}
Instead of $z^\m$(=$Z^\m$) and $w^\m$(=$W^\m$), we take  shifted variables $z'^\m$ and $w'^m$
( As for $Z^0$ and $W^0$, we keep them.), 
and expand $A_\m(z)$ and $A_\m(w)$ around the ``center''
$(x+y)/2$.
\begin{eqnarray}
z=z'+\frac{x+y}{2}\com\nn 
A_\m(z)=A_\m(\frac{x+y}{2})+\pl_\al A_\m|_{\frac{x+y}{2}}\cdot {z'}^\al
+\half\pl_\al\pl_\be A_\m|_{\frac{x+y}{2}}\cdot {z'}^\al{z'}^\be
+O({z'}^3)\ .
\label{DW12}
\end{eqnarray}
The same is for $A_\n(w)=A_\n(w'+\frac{x+y}{2})$.
As a typical calculation example, we show the procedure briefly.
For simplicity we consider the chiral anomaly.
Among terms in (\ref{DW11}), only $\gago\times (\pl A)^2$-terms 
contribute to it.
\begin{eqnarray}
\Tr\,\al(x)\GfMp(X,Y)|_{(\pl A)^2} \sim   \nn
\int^{X^0}_0dZ^0\int^{Z^0}_0dW^0
\int \frac{d^4q}{(2\pi)^4}(-ie^2)
\int d^4x\al(x)\pl_\al A_\m\cdot\pl_\be A_\n   
\times\e^{-iE(q)X^0}
                                    \nn
\times
\tr\left[ \ga_\m \Om_+(q) \ga_\n
\{\frac{\pl\om(q)}{\pl q^\be}
+\Om_+(q) \frac{\pl E(q)}{\pl q^\be}i(-W^0)\}\right.\nn
\times
\left.
\{\frac{\pl\om(q)}{\pl q^\al}
+\Om_+(q) \frac{\pl E(q)}{\pl q^\al}(-i)(X^0-Z^0)\}
\right] \nn
\sim
\half (X^0)^2\int \frac{d^4q}{(2\pi)^4}(-ie^2)
\int d^4x\al(x)\pl_\al A_\m\cdot\pl_\be A_\n   \nn
\times
(\frac{1}{E(q)^4}\ep_{\mn\al\tau}q^\tau q^\be
+\frac{1}{2E(q)^2}\ep_{\mn\be\al})\e^{-iE(q)X^0}
\com
\label{DW12x}
\end{eqnarray}
where $\om(k)\equiv \frac{M+i\kslash}{2E(k)}$.
The notation "$\sim$", here and in the following, means "equal up to irrelevant terms".
We do the momentum ($q^\m$) integral in $t$-coordinate 
(not in $X^0$-coordinate, $X^0=-it$). 
The final result is obtained by taking the limit
$Mt\ra +0$.
\begin{eqnarray}
\Tr\,\al(x)\GfMp(X,Y)|_{(\pl A)^2}
\sim
-\frac{t^2}{2}(-ie^2)
\int d^4x\al(x)\pl_\al A_\m\cdot\pl_\be A_\n   \nn
\times
\frac{M^2}{8\pi^2}[ \ep_{\mn\al\tau}\frac{\del_{\tau\be}}{4}
E^5_4(Mt)
+\half\ep_{\mn\be\al}E^3_2(Mt)]\nn
\ra
\frac{ie^2}{64\pi^2\times 4}\int d^4x\al(x)F_\ab{\tilde F}_\ab\com\q
Mt\ra +0\com \nn
\label{DW12y}
\end{eqnarray}
where 
the function $E^r_n(a)$ is defined in App.B and the results $a^2E^5_4(a)\ra 1,
a^2E^3_2(a)\ra 1$ are used (see (\ref{Int.4})).
In the above evaluation in $t$-coordinate, the momentum ($q^\m$) integral
is convergent. 
This is because we take only positive (negative) energy states
for (+)-domain ((-)-domain). 
This is one advantageous point of the Feynman path.
We can obtain the same result in the $X^0$-coordinate
\cite{SI98}. In this case, however, 
the momentum integral must be regularized and we take
the analytic continuation. 
(This is always necessary for the anti-Feynman and symmetric path.) 
Since it is one of important
points of the present regularization, 
we explain it separately in the next section.  

$\GfMm|_{AA}$ is similarly calculated (see Fig.4(i)).
\begin{eqnarray}
X^0<Z^0<W^0<Y^0=0\nn
\GfMm|_{AA}
=\int^0_{X^0}dZ^0\int_{Z^0}^0dW^0\int d^4Z\int d^4W   \nn
\times G^n_0(X,Z)i\e \Aslash(z)G^n_0(Z,W)
i\e \Aslash(w)G^n_0(W,Y)\nn
=\int_{X^0}^0dZ^0\int_{Z^0}^0dW^0
\int d^4z\int d^4w\int\frac{d^4k}{(2\pi)^4}\frac{d^4l}{(2\pi)^4}
\frac{d^4q}{(2\pi)^4}\nn
\times (-i)\Om_-(k) i e \Aslash(z)
(-i)\Om_-(l)
i e \Aslash(w)(-i)\Om_-(q)\nn
\times\exp \{i\Kbar(X-Z)+i\Lbar(Z-W)+i\Qbar(W-Y)\}
\com
\label{DW12b}
\end{eqnarray}
where $\Om_-(k)\equiv\frac{M-\Kbarbslash}{2E(k)}$.
Using this expression, $\Tr\,\al(x)\GfMm(X,Y)|_{(\pl A)^2}$ 
turns out to be the same as (\ref{DW12y})
except the sign. 
\begin{figure}
\centerline{\epsfysize=6cm\epsfbox{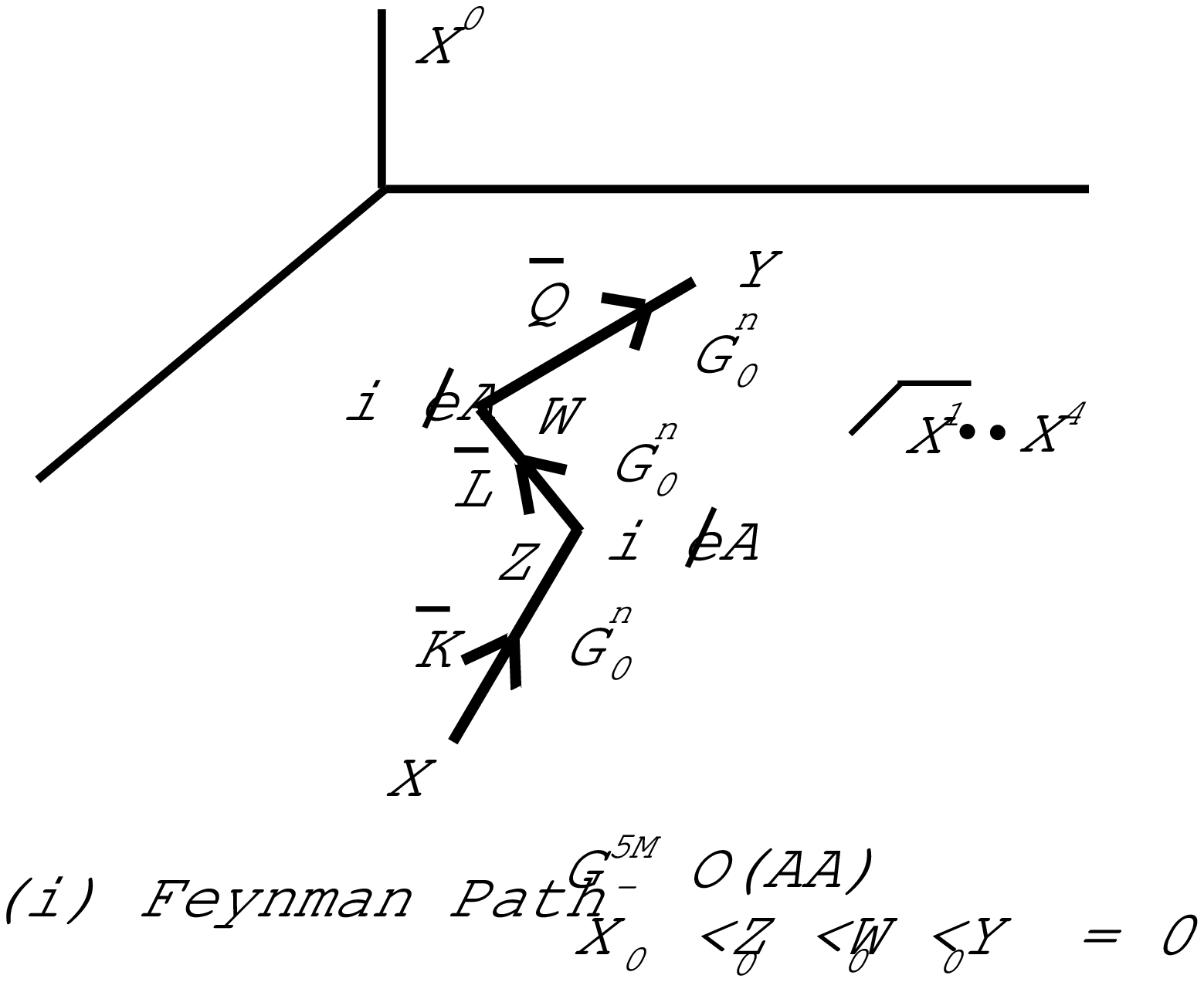}}
   \begin{center}
Fig.6\ 
Abelian Gauge Theory, $\GfMm$, $O(AA)$, Feynman Path.
   \end{center}
\end{figure}

Finally we obtain the ABJ anomaly as {\it one fourth} of (\ref{QED9}).
\footnote{
The discrepant factor $1/4$ comes from the fact that the Feynman path
is not the proper solution as commented in Subsec.3.1. If we allowed
to consider in the extreme chiral limit, hoping that the chiral
anomaly itself is a topological object and does not depend on the
continuous parameter $M$, we can explain the factor as follows. 
For each vertex, 
instead of $\Aslash=\frac{1+\gago}{2}\Aslash+\frac{1-\gago}{2}\Aslash$,
we consider "half" of it ( the right-part for Feynman ). Therefore
$(\half)^2$ factor appears for $O(A^2)$ contribution.
}


\subsection{Symmetric Path}
We sketch the derivation of ABJ anomaly of 4 dim QED
using the symmetric path. 
The first eq. of (\ref{DW13bb}) is evaluated using 
the following expression(see Fig.5(ii)).
\begin{eqnarray}
\GfMp|_{AA}
=\int^{X^0}_0dZ^0\int^{Z^0}_0dW^0\int d^4Z\int d^4W
(G^p_0(X,Z)-G^n_0(X,Z))ie\Aslash(z)  \nn
(G^p_0(Z,W)-G^n_0(Z,W))ie\Aslash(w)
(G^p_0(W,Y)-G^n_0(W,Y))             \nn
=\int^{X^0}_0dZ^0\int^{Z^0}_0dW^0\int d^4z\int d^4w   
\int\frac{d^4k}{(2\pi)^4}\frac{d^4l}{(2\pi)^4}\frac{d^4q}{(2\pi)^4}\nn
\times (-i)(\Om_+(k)\e^{-iE(k)(X^0-Z^0)}
           -\Om_-(k)\e^{iE(k)(X^0-Z^0)})
ie\Aslash(z)                             \nn
\times (-i)(\Om_+(l)\e^{-iE(l)(Z^0-W^0)}
           -\Om_-(L)\e^{iE(l)(Z^0-W^0)})
ie\Aslash(w)                             \nn
\times (-i)(\Om_+(q)\e^{-iE(q)(W^0-Y^0)}
           -\Om_-(q)\e^{iE(q)(W^0-Y^0)})
\e^{-ik(x-z)-il(z-w)-iq(w-y)}
\pr
\label{DW13c}
\end{eqnarray}
We will realize the properties of $\Om_\pm$
presented in App.C efficiently works here. We evaluate
the measure change (\ref{DW13bb}):\ 
$\del_\al\,\ln\,J_{ABJ}=
\lim \Tr\,i^2\al(x)(\GfMp(X,Y)-\GfMm(X,Y))$.
For later use we call the following procedure "Standard Calculation
(SC) Procedure". 
\nl
1)\ expanding $A_\m(z)$ and $A_\n(w)$ around the "center"
$(x+y)/2$,\nl 
2)\ integrating out
the shifted coordinates:\ 
$w'=w-\frac{x+y}{2},\ z'=z-\frac{x+y}{2}$\ ,\nl
3)\ integrating out the momenta $q_\m, l_\m$,\nl 
4)\ partial integrations,\nl
5)\ taking the trace for 4 dim space coordinate x=y,\nl
After applying the above procedure to $\Tr\,\al(x)\GfMp(X,Y)$ using
the expression (\ref{DW13c}), 
its relevant part is given by
\begin{eqnarray}
\Tr \al(x)\GfMp(X,Y)|_{(\pl A)^2}   
=\int^{X^0}_0dZ^0\int^{Z^0}_0dW^0\int d^4x   
\al(x)(-ie^2)\pl_\al A_\m\cdot\pl_\be A_\n \nn
\times 
\int\frac{d^4k}{(2\pi)^4}\times F_{\al\m\be\n}(k;X^0,Z^0,W^0)\com\nn
F_{\al\m\be\n}\equiv\tr
\{
\pl_\be (\Omp\e^{-iEW^0}-\Omm\e^{iEW^0})\cdot
\pl_\al (\Omp\e^{-iE(X^0-Z^0)}-\Omm\e^{iE(X^0-Z^0)})\cdot\ga_\m\nn
\times (\Omp\e^{-iE(Z^0-W^0)}-\Omm\e^{iE(Z^0-W^0)})\cdot\ga_\n\}
\com
\label{ABJ1}
\end{eqnarray}
where 
$\pl_\al=\frac{\pl}{\pl k^\al},\ \Om_\pm=\Om_\pm(k),\ 
E=E(k),\ $ and $Y^0=0$.
Here we focus only on $(\pl A)^2$-part because 
it is sufficient for the ABJ anomaly. 
Now we use
the following relations:
\begin{eqnarray}
\pl_\al (\Omp\e^{-iE(X^0-Z^0)})
=\{ \pl_\al\om-i\pl_\al E\cdot (X^0-Z^0)\Omp\}\e^{-iE(X^0-Z^0)}\ ,\nn 
\om=\om(k)=\frac{M+i\kslash}{2E}\ ,\nn
\Omp\Omp=\frac{M}{E}\Omp\com\q \Omm\Omm=\frac{M}{E}\Omm\com\nn
\Omp\Omm=\frac{M}{E}\Omp-\gago\Omp(-k)\com\q
\Omm\Omp=\frac{M}{E}\Omm+\gago\Omm(-k)\pr
\label{ABJ2}
\end{eqnarray}
The full list of useful relations involving $\Omp,\Omm$ are
given in App.C. 
Especially the projective property between $\Omp$ and $(\Omm)^\dag$
and between $\Omm$ and $(\Omp)^\dag$ should be noted.

$F_{\al\m\be\n}$ is rewritten as
\begin{eqnarray}
F_{\al\m\be\n}=\tr \left[
\pl_\be\om\cdot\pl_\al\om(\ga_\m\Omp\ga_\n\e^{-iEX^0}
                         -\ga_\m\Omm\ga_\n\e^{iEX^0})
						 \right.\nn
-\{\pl_\be\om\cdot\pl_\al\om-i\pl_\be\om\cdot\pl_\al E\cdot (X^0-Z^0)\Omp
-i\pl_\be E\cdot W^0\Omp\pl_\al\om\}   \nn
\times\ga_\m\Omm\ga_\n\e^{-iE(X^0-2Z^0+2W^0)}\nn
+\{\pl_\be\om\cdot\pl_\al\om+i\pl_\be\om\cdot\pl_\al E\cdot (X^0-Z^0)\Omm
+i\pl_\be E\cdot W^0\Omm\pl_\al\om\}  \nn
\times\ga_\m\Omp\ga_\n\e^{iE(X^0-2Z^0+2W^0)}\nn
-\{\pl_\be\om\cdot\pl_\al\om+i\pl_\al E\cdot (X^0-Z^0)\pl_\be\om\cdot\Omm
\} \ga_\m\Omp\ga_\n\e^{iE(X^0-2Z^0)}\nn
+\{\pl_\be\om\cdot\pl_\al\om-i\pl_\al E\cdot (X^0-Z^0)\pl_\be\om\cdot\Omp
\} \ga_\m\Omm\ga_\n\e^{-iE(X^0-2Z^0)}\nn
+\{\pl_\be\om\cdot\pl_\al\om-i\pl_\be E\cdot W^0\Omp\pl_\al\om
\} \ga_\m\Omm\ga_\n\e^{iE(X^0-2W^0)}\nn
                   \left.
-\{\pl_\be\om\cdot\pl_\al\om+i\pl_\be E\cdot W^0\Omm\pl_\al\om
\} \ga_\m\Omp\ga_\n\e^{-iE(X^0-2W^0)}
                   \right]
\com
\label{ABJ3}
\end{eqnarray}
where the following relations are used to eliminate some terms. 
As for the contribution to the chiral anomaly, 
we can confirm
\begin{eqnarray}
\tr \Om_\pm\ga_\m\Om_\pm\ga_\n\sim 0\q (\mbox{arbitrary choice for}\ \pm)\com\nn
\tr \gago\Om_\pm(-k)\ga_\m\Om_\pm(k)\ga_\n\sim 0\q (\mbox{arbitrary choice for}\ \pm)\com\nn
\tr \pl_\al\om\cdot\Omp\ga_\m\Omp\ga_\n\sim 0\com\q
\tr \pl_\al\om\cdot\Omm\ga_\m\Omm\ga_\n\sim 0\com\nn
\tr \Omp\pl_\al\om\cdot\ga_\m\Omp\ga_\n\sim 0\com\q
\tr \Omm\pl_\al\om\cdot\ga_\m\Omm\ga_\n\sim 0\com
\label{ABJ4}
\end{eqnarray}
where all other terms than the $\ep$-tensor term are ignored
in the right-hand sides of above equations.
Further evaluation goes with the help of the
following useful relations valid for the chiral anomaly ($\ep$-tensor) part.
\begin{eqnarray}
\tr \pl_\al\om\cdot\Omp\ga_\m\Omm\ga_\n\sim +\frac{k^\tau}{E^2}\ep_{\tau\al\mn}\com\q
\tr \pl_\al\om\cdot\Omm\ga_\m\Omp\ga_\n\sim -\frac{k^\tau}{E^2}\ep_{\tau\al\mn}\com\nn
\tr \Omp\pl_\al\om\cdot\ga_\m\Omm\ga_\n\sim -\frac{k^\tau}{E^2}\ep_{\tau\al\mn}\com\q
\tr \Omm\pl_\al\om\cdot\ga_\m\Omp\ga_\n\sim +\frac{k^\tau}{E^2}\ep_{\tau\al\mn}\com\nn
\tr \pl_\be\om\cdot\pl_\al\om\cdot\ga_\m\Om_\pm\ga_\n\sim 
\mp\frac{1}{2}(\frac{1}{E^2}-\frac{k^2}{2E^4})\ep_{\n\be\al\m}\ 
(\mbox{corresponding choice for}\ \pm)\pr
\label{ABJ4b}
\end{eqnarray}

Finally  $\Tr \al(x)\GfMp$ reduces to
\begin{eqnarray}
\Tr \al(x)\GfMp(X,Y)|_{(\pl A)^2}   
\sim\nn
\int d^4x \al(x)(-ie^2)\pl_\al A_\m\cdot\pl_\be A_\n
\int^{X^0}_0dZ^0\int^{Z^0}_0dW^0     \nn
\times\int\frac{d^4k}{(2\pi)^4} \left[
\half (\frac{1}{E^2}-\frac{k^2}{2E^4})\ep_{\al\m\be\n}\times
2\{ \cos EX^0+\cos E(X^0-2Z^0+2W^0)
 \right.
\nn
-\cos E(X^0-2Z^0)-\cos E(X^0-2W^0) \}  \nn
+i\frac{k^\tau}{E^2}  
      \left[ \{ (X^0-Z^0)\pl_\al E\cdot\ep_{\tau\be\mn}
-W^0\pl_\be E\cdot\ep_{\tau\al\mn} \}(-2i)\sin E(X^0-2Z^0+2W^0) 
 \right.
\nn
 \left.
       \left. 
+(X^0-Z^0)\pl_\al E\cdot\ep_{\tau\be\mn}2i\sin E(X^0-2Z^0)
+W^0\pl_\be E\cdot\ep_{\tau\al\mn}2i\sin E(X^0-2Z^0)
       \right] 
\right]  \nn
\mbox{}\nn  
=\int d^4x \al(x)(-ie^2)\pl_\al A_\m\cdot\pl_\be A_\n
\int\frac{d^4k}{(2\pi)^4}\times   \nn
\left[
\half (\frac{1}{E^2}-\frac{k^2}{2E^4})\ep_{\be\n\al\m}
((X^0)^2\cos EX^0-\frac{X^0}{E}\sin EX^0)              
\right.
\nn
\left.
+\frac{1}{2E^5}(-E(X^0)^2\cos EX^0+X^0\sin EX^0)\times
                     \fourth k^2\ep_{\ab\mn}\times 2 
\right] \nn
\mbox{}\nn  
=\int d^4x \al(x)(-ie^2)\pl_\al A_\m\cdot\pl_\be A_\n\ep_{\be\n\al\m}\times
\frac{2\pi^2}{(2\pi)^4}  \nn
\times\{ \half (MX^0)^2(C^3_2(MX^0)-\half C^5_4(MX^0))
        -\half MX^0 (S^3_3(MX^0)-\half S^5_5(MX^0))\nn
+\fourth (MX^0)^2C^5_4(MX^0)-\fourth MX^0S^5_5(MX^0)\}\nn
\ra \int d^4x \al(x)(-ie^2)\frac{2\pi^2}{(2\pi)^4}
\{ \half ( (-1)-\half (-1) )   \nn
-\half (1-\half\times 1)
+\fourth (-1)-\fourth\times 1 \}\fourth  
F_\mn \Ftil_\mn  \nn
=\int d^4x \al(x)(ie^2)\frac{1}{16\pi^2}\half F_\mn \Ftil_\mn
\q\q \mbox{as}\q\q MX^0\ra +0
\com
\label{ABJ5}
\end{eqnarray}
where $S^r_n(a)$ and $C^r_n(a)$ are defined in App.B. Adding the (-)-domain
contribution $\Tr \al(x)\GfMm(X,Y)$, which is the same as above except
the sign, we obtain the correct value of ABJ anomaly (\ref{QED9}).

\vspace{1cm}

\q The Weyl anomaly can be evaluated using the second equation
of (\ref{DW10}) (Feynman) or (\ref{DW13bb}) (Symmetric)
in the similar way above. In this case   
we need to consider the parity-even terms instead of the odd ones
($\gago$ terms) in the trace. We also need to take into account 
$A\pl\pl A$-terms besides $(\pl A)^2$-terms.
It is one of the present advantage that both Weyl and chiral anomalies can be
treated in a common framework. Details have been found
in \cite{SI99aei}.

\section{ 
Regularization of Momentum Integral }
As shown in the cut-off parameter
$M$ (\ref{DW8a}), the regularization in the momentum integral
is one of most important points in the present approach.
On the one hand the explanation about the chiral condition can be
clearly stated in terms of the cutoff parameter $M$ as we have done in Sect.3.
From this viewpoint, the present regularization is explained in (i) part of the
following. On the other hand we {\it cannot} explicitly introduce the cutoff parameter
$M$ in the momentum integral because it breaks the gauge invariance. The present
regularization is re-examined in (ii) from this viewpoint. ( Another
possible regularization is the higher-derivative one, which corresponds
to the Wilson term as commented in a footnote of (\ref{DW8a}). )
We take two characteristic momentum integrals
which are divergent:\  
\begin{eqnarray}
F_{s}(a)\equiv
\int^\infty_0 dx\frac{x(x^2+2)}{(\sqxx)^3}\sin (a\sqxx)
=\int^\infty_1 dy\frac{(y^2+1)}{y^2}\sin (ay)\ ,\nn
F_{c}(a)\equiv
\int^\infty_0 dx\frac{x(x^2+2)}{(\sqxx)^3}\cos (a\sqxx)
=\int^\infty_1 dy\frac{(y^2+1)}{y^2}\cos (ay)\ ,
\label{MI.1}
\end{eqnarray}
where $a>0\ ,\ y=\sqxx$ and $x$ appears, in the concrete calculation, as
$x=\sqrt{k^\m k^\m}/M$\ ($k^\m$\ :\ momentum in the 4 dim
Euclidean space;\ $M$\ :\ 1+4 dim fermion mass). 
In terms of the notations in App.B, these two integrals are
$F_s(a)=S^3_3(a)+2S^1_3(a),\ F_c(a)=C^3_3(a)+2C^1_3(a)$.

\flushleft{(i)\ Use of exponetially damping factor}

\q We can {\it regularize} above ones as
\begin{eqnarray}
F_{c}(a)+iF_{s}(a)
=\lim_{\ep\ra +0}\int^\infty_1 dy\frac{(y^2+1)}{y^2}\e^{i(a+i\ep)y}\nn
=-\frac{\sin a}{a}+\cos a-a\int_a^\infty\frac{\sin y}{y}dy\nn
+i\{ \frac{\cos a}{a}+\sin a+a\int_a^\infty\frac{\cos y}{y}dy\}\com
\label{MI.2}
\end{eqnarray}
where $\ep\ (\ra +0)$ is a positive regularization parameter
for the convergence.
As for the correspodence with the chiral condition (\ref{DW8a}), 
the above regularization is essentially
equivelent to taking the following condition.
\begin{eqnarray}
x=\frac{\sqrt{k^2}}{M}\sim y\leq 1\pr
\label{MI.2b}
\end{eqnarray}
We use some formulae
\begin{eqnarray}
\mbox{si} (a)\equiv -\int^\infty_a\frac{\sin y}{y} dy
=\mbox{Si} (a)-\frac{\pi}{2}\com                                 \nn
\mbox{Si} (a)\equiv \int^a_0\frac{\sin y}{y} dy
=\sum_{n=0}^{\infty}\frac{(-1)^na^{2n+1}}{(2n+1)!(2n+1)}\com   \nn
\mbox{ci}(a)\equiv\mbox{Ci}(a)\equiv -\int^\infty_a\frac{\cos y}{y} dy
=\ga+\ln\,a+\int^a_0\frac{\cos\,y-1}{y}dy\com                  \nn
\cos a+ a\,\mbox{Si} (a)=\mbox{}_1F_2(-\half;\half,\half;-\frac{a^2}{4})\com\nn
\sin a-a\int^a_0\frac{\cos\,y-1}{y}dy=a+\frac{a^3}{12}\,
\mbox{}_2F_3(1,1;2,2,\frac{5}{2};-\frac{a^2}{4})\com
\label{MI.3}
\end{eqnarray}
where si($a$), Si($a$), ci($a$) and Ci($a$) are
integral functions and $\mbox{}_pF_q$ is the generalized hypergeometric
function (see App.B). The appearance of those functions clearly distinguishes
the present regularization from other ones (dimensional, usual Pauli-Villars, etc.).
We finally obtain exact expressions.
\begin{eqnarray}
F_{s}(a)=a-a\ga+\frac{\cos a}{a}-a\ln\,a
+\frac{a^3}{12}
\mbox{}_2F_3(1,1;2,2,\frac{5}{2};-\frac{a^2}{4})\com\nn
F_{c}(a)=-\frac{\pi}{2}a-\frac{\sin a}{a}
+\mbox{}_1F_2(-\half;\half,\half;-\frac{a^2}{4})\com
\label{MI.4}
\end{eqnarray}
where  $\mbox{}_2F_3$ and $\mbox{}_1F_2$ are regular
at $a\ra +0$ and has the following forms:
\begin{eqnarray}
\mbox{}_1F_2(-\half;\half,\half;-\frac{a^2}{4})=
1-\half\sum_{n=1}^{\infty}\frac{(-1)^na^{2n}}{(2n-1)!(2n-1)n}
=1+\half a^2+O(a^4)\ ,\nn
\mbox{}_2F_3(1,1;2,2,\frac{5}{2};-\frac{a^2}{4})=
6\sum_{n=0}^{\infty}\frac{(-1)^na^{2n}}{(2n+3)!(n+1)}
=1-\frac{1}{40}a^2+O(a^4)\pr
\label{MI.5}
\end{eqnarray}
Therefore we have the following limit:
\begin{eqnarray}
a\,F_s(a)\ra 1\com\q
F_c(a)\ra 0\com\q \mbox{as}\q a\ra +0\pr 
\label{MI.5b}
\end{eqnarray}

\flushleft{(ii)\ Analytical Continuation}

\q Instead of (i) we can do the same thing by the 
{\it analytic continuation}, which donot need to
 introduce an additional regularization parameter.
We start with a convergent integral :
\begin{eqnarray}
F_e(a)\equiv
\int^\infty_0 dx\frac{x(x^2+2)}{(\sqxx)^3}\e^{-a\sqxx}\nn
=\frac{\e^{-a}}{a}+\e^{-a}+a(\ga+\ln\,a+
\sum_{n=1}^{\infty}\frac{(-1)^na^n}{n!n})\com\q a>0
\com
\label{MI.6}
\end{eqnarray}
where $F_e(a)=E^3_3(a)+2E^1_3(a)$. Let us define $F_c(a)$
and $F_s(a)$ by the following analytic continuation.
\begin{eqnarray}
a\ra -ia \com\q
F_e(a)\ra F_e(-ia)=F_{c}(a)+iF_{s}(a)\pr
\label{MI.6b}
\end{eqnarray}
In this case we must specify a branch $N=-1$ in
$\ln(-ia)=(\frac{3}{2}+2N)\pi i+\ln\,a$ in order
to obtain the results (\ref{MI.4}). 
Note that the final limit ($a\ra +0$) is not affected by this
ambiguity $N$.

\vspace{1cm}

\q In Fig.7, we plot the integrand of $F_{s}(a)$
of (\ref{MI.1}) and its regularized one:
\begin{eqnarray}
f(x,a;\ep)
=\frac{x(x^2+2)}{(\sqxx)^3}\sin (a\sqxx)\e^{-\ep\sqxx}\com\q
a=10.0\com\q \ep=1.0\pr
\label{MI.7}
\end{eqnarray}
\begin{figure}
\centerline{\epsfysize=4cm\epsfbox{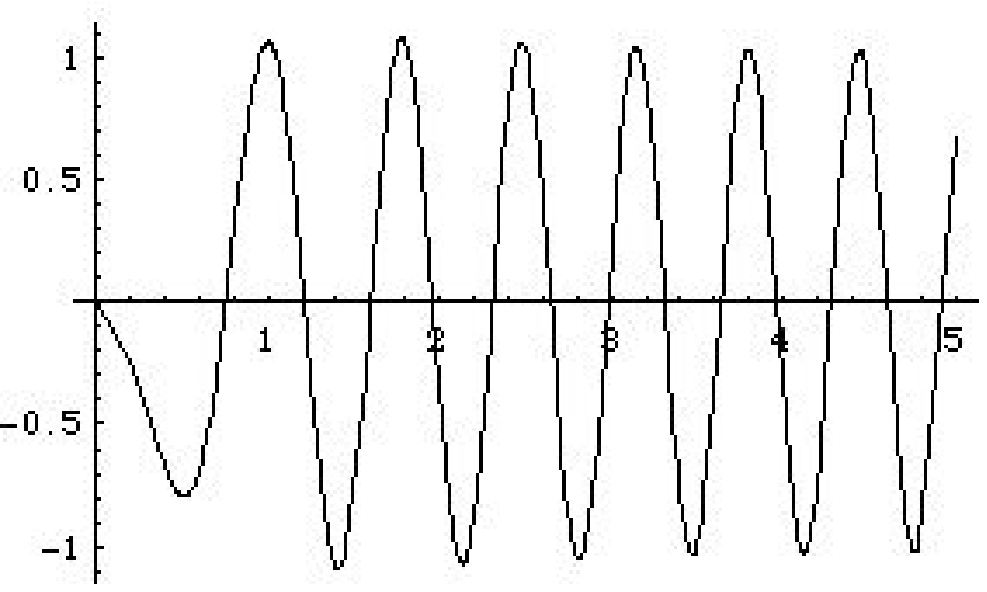}\\
            \epsfysize=4cm\epsfbox{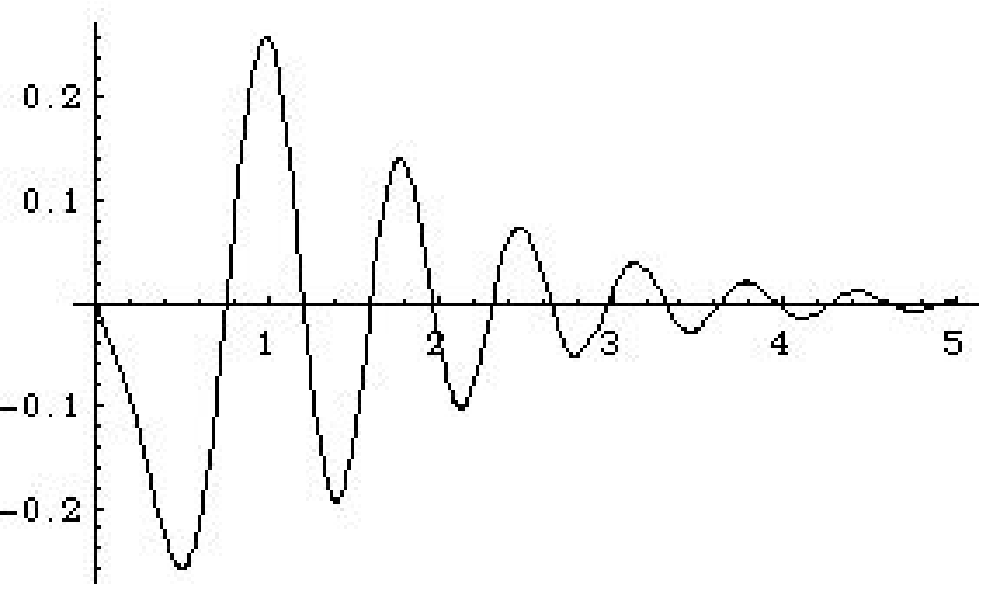}}
   \begin{center}
Fig.7\ 
Graphs of 
$f(x,a=10.0;\ep=0.0)
=\frac{x(x^2+2)}{(\sqxx)^3}\sin (10.0\sqxx)\ $
(above) and its regularized one 
$f(x,a=10.0;\ep=1.0)
=\frac{x(x^2+2)}{(\sqxx)^3}\sin (10.0\sqxx)\e^{-\sqxx}\ $
(below). $0\leq x\leq 5$. See the integrand of (\ref{MI.1}) and 
its regularized one (\ref{MI.7}).
   \end{center}
\end{figure}
The present regularization, in the momentum integral, typically do the
following things:\ 
1) (Exponetially) divergent functions, due to negative
eigenvalues, are first replaced by oscillating ones
(Fig.7, above) using the Wick rotation ($X^0=\mp it$) and then
2) the large momentum ($|k^\m|\geq M$) region is made
to be exponentially damped (Fig.7, below) 
by the $\ep$-factor (or the analytic continuation, or the chiral condition
(\ref{DW8a})). 
Finally we take 
3) the limit $Mt\ra +0$ or $MX^0\ra \pm 0$.
Steps 1) and 2) are for ultra-violet regularization whereas
3) is for infrared one.

\section{2 Dim Non-Abelian Anomaly}

Let us consider 2 dim chiral non-Abelian gauge theory and
analyze its anomaly in the present domain wall approach.
Its consistent and covariant anomalies were examined
in the ordinary heat-kernel by \cite{Leut85,DAMTP9687} and 
in the ordinary domain wall approach by \cite{NN94,DS97PL,Neub98b}. 
\begin{eqnarray}
\Lcal=\psibar\Dhat\psi\com\nn
\Dhat=i\ga_\m(\pl_\m+iP_+ R_\m)\com\q
\Dhat^\dag=i\ga_\m(\pl_\m+iP_- R_\m)  \com\nn
P_\pm=\half (1\pm\gago)   \com
\label{NA1}
\end{eqnarray}
where $R_\m$ is the chiral (right-handed) gauge field.
$\Dhat$ is {\it not} hermitian, which is a different point from 
the (4 dim) QED of Sec.3. 
The lagrangian has the chiral gauge symmetry.
\begin{eqnarray}
\psi'=\e^{iP_+\la(x)}\psi\com\q
\psibar'=\psibar\e^{-iP_-\la(x)}\com\q \la(x)=T^a\la^a(x)\com\nn
{R_\m}'=U(\la)R_\m U^{-1}(\la)+i\pl_\m U(\la)\cdot U^{-1}(\la)\com\q
U(\la)=\e^{i\la(x)}  \com
\label{NA2}
\end{eqnarray}
where $\la(x)$ is the gauge parameter and $T^a$ is the generators
of the symmetry group. In the above notation the field strength
and its transformation is given by
$F_\mn=\pl_\m R_\n-\pl_\n R_\m+i[R_\m,R_\n]\ ,\ 
{F_\mn}'=U(\la)F_\mn U^{-1}(\la)$.  
The variation of the
Jacobian for the change of variables 
($\psi,\psibar \ra \psi',\psibar'$) is given by
\begin{eqnarray}
\del_\la\,\ln\, J_{NA}=
\del_\la\,\ln\,\left|\frac{\pl (\psi',\psibar')}{\pl (\psi,\psibar)}\right|
                                   \nn
=
\left\{ \begin{array}{c}
\Tr\,i\la(x)\,(P_+\del^2(x-y)-P_-{\bar {\del^2}}(x-y)\,)+O(\la^2)\com \\
                        \mbox{or}\\
\Tr\,i\la(x)\,\gago\del^2(x-y)+O(\la^2)\com\\
                        \mbox{or}\\
\half\Tr\,i\la(x)\,(\gago\del^2(x-y)+\gago{\bar {\del^2}}(x-y)\,)
+O(\la^2)\com
\end{array} \right.
\pr
\label{NA3}
\end{eqnarray}
We have different choices here.
${\bar {\del^2}}(x-y)$ stresses that its regularization form is not
necessarily the same as that of  $\del^2(x-y)$ at the intermediate stage.
Because both $\Dhat$ and $\Dhat^\dag$ satisfy the relation (\ref{DW3}),
we have
\begin{eqnarray}
\ln\,Z[R]=\ln\,
\int\Dcal\psibar\Dcal\psi\e^{-\intfx\Lcal}
=\Tr\ln\Dvec=-\Tr\int_0^\infty\frac{\e^{-t\Dvec}}{t}dt+\mbox{const}\nn
=-\int_0^\infty\frac{dt}{t}\Tr 
[\half(1+i\ga_5)\e^{+it\ga_5\Dvec}+
\half(1-i\ga_5)\e^{-it\ga_5\Dvec}]+\mbox{const}\ ,\nn
(\ln\,Z[R])^*
=\Tr\ln\Dvec^\dag=-\Tr\int_0^\infty\frac{\e^{-t\Dvec^\dag}}{t}dt
+\mbox{const}\nn
=-\int_0^\infty\frac{dt}{t}\Tr 
[\half(1+i\ga_5)\e^{+it\ga_5\Dvec^\dag}+
\half(1-i\ga_5)\e^{-it\ga_5\Dvec^\dag}]+\mbox{const}\ .
\label{NA4}
\end{eqnarray}
Some heat-kernels are naturally introduced.
\begin{eqnarray}
G^{5M}_\pm (x,y;t)\equiv <x|\exp\{\pm it\ga_5(\Dvec+iM)\}|y>\com\nn
G^{5M}_{c\pm} (x,y;t)\equiv 
<x|\exp\{\pm it\ga_5(\Dvec^\dag+iM)\}|y>\com\nn
G^{5M}_{h\pm} (x,y;t)\equiv 
<x|\exp\{\pm it\ga_5(\Dvec_h+iM)\}|y>\com\nn
\Dhat_h=i\ga_\m(\pl_\m+iR_\m)\com\q
\Dhat_h=(\Dhat_h)^\dag\pr
\label{NA5}
\end{eqnarray}
\footnote{
We should not take 
$\half(\Dhat+\Dhat^\dag)=i\ga_\m(\pl_\m+\frac{i}{2}R_\m)$
as the hermitian operator $\Dhat_h$ because, in this case,
$R_\m$ transforms as
${R_\m}'/2=U(\la)(R_\m/2)U^{-1}(\la)+i\pl_\m U(\la)\cdot U^{-1}(\la)$,
not as in (\ref{NA2}).
}
The above heat-kernels satisfy the 1+2 dim
Minkowski Dirac equation after appropriate Wick rotations for $t$.
\begin{eqnarray}
(i\dbslash-M) G^{5M}_\pm =i \Rslash P_+ G^{5M}_\pm\com\q
(X^a)=(\mp it,x^\m)\com                    \nn
(i\dbslash-M) G^{5M}_{c\pm} =i \Rslash P_- G^{5M}_{c\pm}\com\q
(X^a)=(\mp it,x^\m)           \com\nn
(i\dbslash-M) G^{5M}_{h\pm} =i \Rslash G^{5M}_{h\pm}\com\q
(X^a)=(\mp it,x^\m)           \pr
\label{NA6}
\end{eqnarray}
We notice the existence of different choices for the present regularization.
See Fig.8. 
Among them we consider two representative ones which  correspond to
the consistent and covariant anomalies.

\begin{figure}
\centerline{\epsfysize=4cm\epsfbox{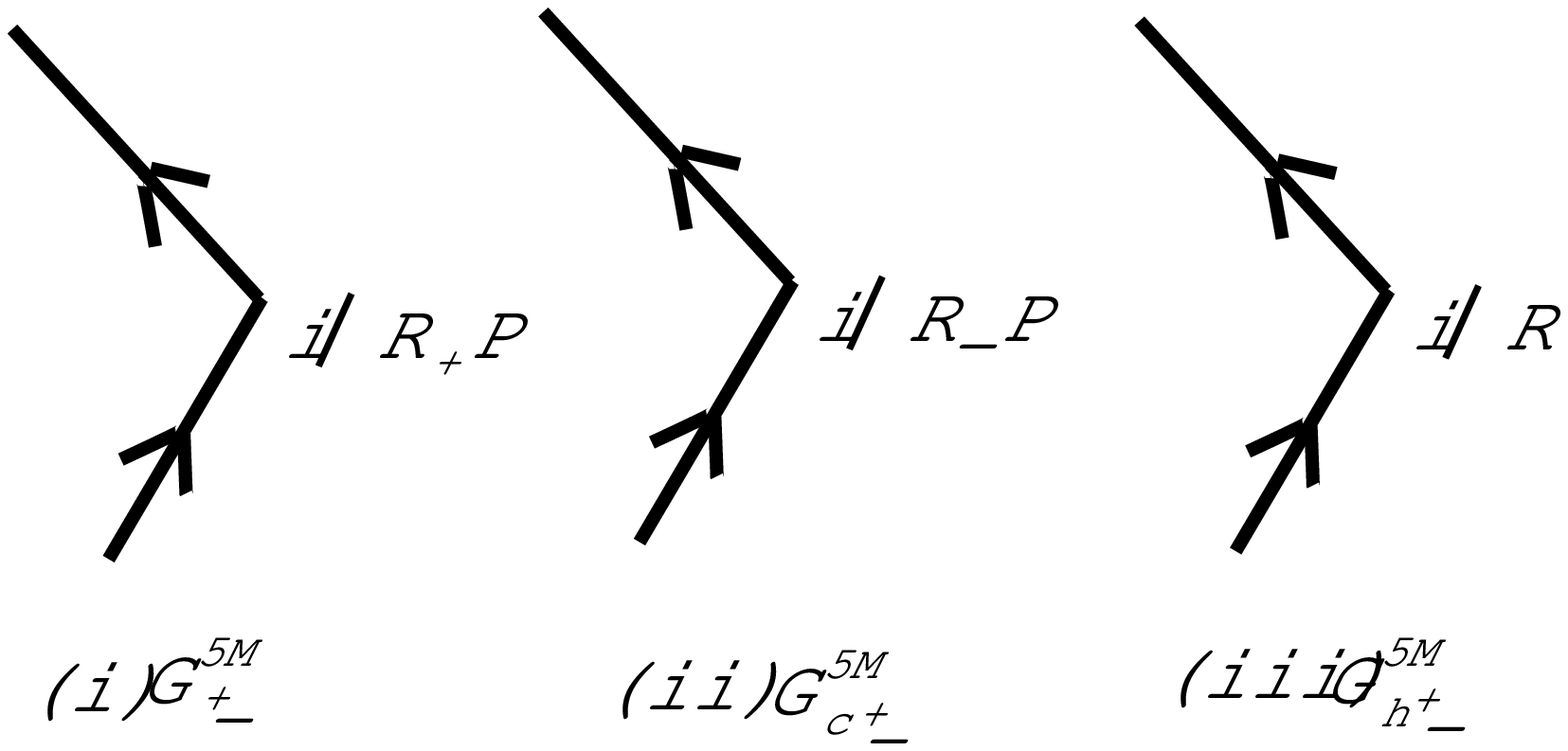}}
   \begin{center}
Fig.8\ Vertices corresponding to $G^{5M}_\pm,G^{5M}_{c\pm},G^{5M}_{h\pm}$
which are defined by (\ref{NA6}). 
   \end{center}
\end{figure}

\flushleft{(i) Consistent Anomaly}

Let us first take the Feynman path and $G^{5M}_\pm$. 
From the middle eq. of
(\ref{NA3}) and the boundary condition of Feynman path
solution (\ref{DW9c}), we obtain
\begin{eqnarray}
\del_\la\,\ln\, J_{NA}
=\lim_{Mt\ra +0}\Tr\,i\la(x)\,
(i \GfMp(x,y;t)-i \GfMm(x,y;t)\ )+O(\la^2)\pr
\label{NA6x}
\end{eqnarray}
We consider the 1-st and 2-nd order (with respect to $R_\m$) perturbations.
The first term of (\ref{NA6x}) is evaluated from
\begin{eqnarray}
\GfMp|_R
=\int^{X^0}_0dZ^0\int d^2ZG^p_0(X,Z)i \Rslash(z)P_+G^p_0(Z,Y)\nn
=\int^{X^0}_0dZ^0\int d^2z\int\frac{d^2k}{(2\pi)^2}\frac{d^2l}{(2\pi)^2}\nn
\times (-i)\Omp(k) i \Rslash(z)P_+
(-i)\Omp(l)
\e^{-i\Ktil(X-Z)}\e^{-i\Ltil(Z-Y)}
\pr
\label{NA7}
\end{eqnarray}
See Fig.9(i).
\begin{figure}
\centerline{\epsfysize=6cm\epsfbox{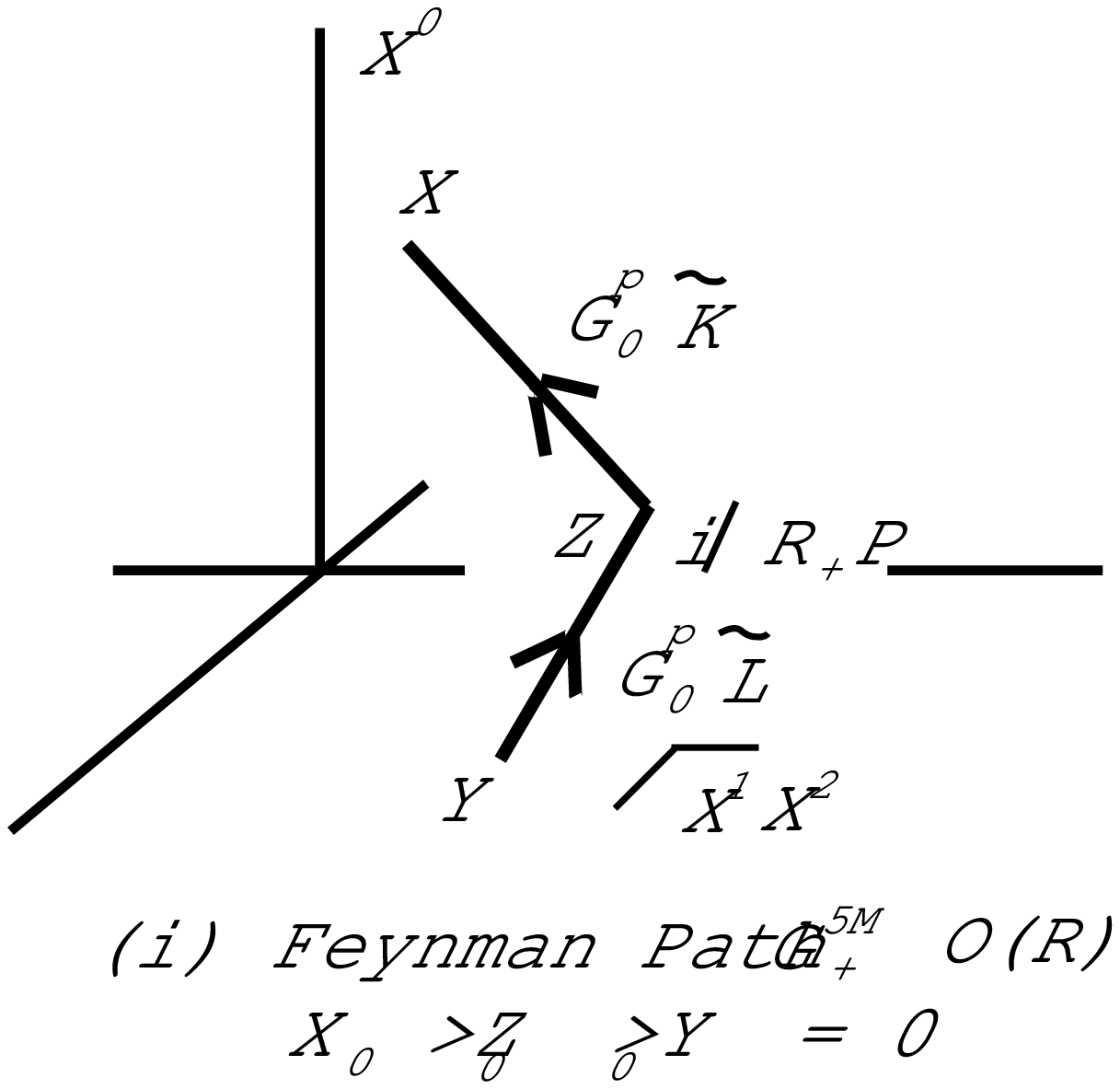}\ \
            \epsfysize=6cm\epsfbox{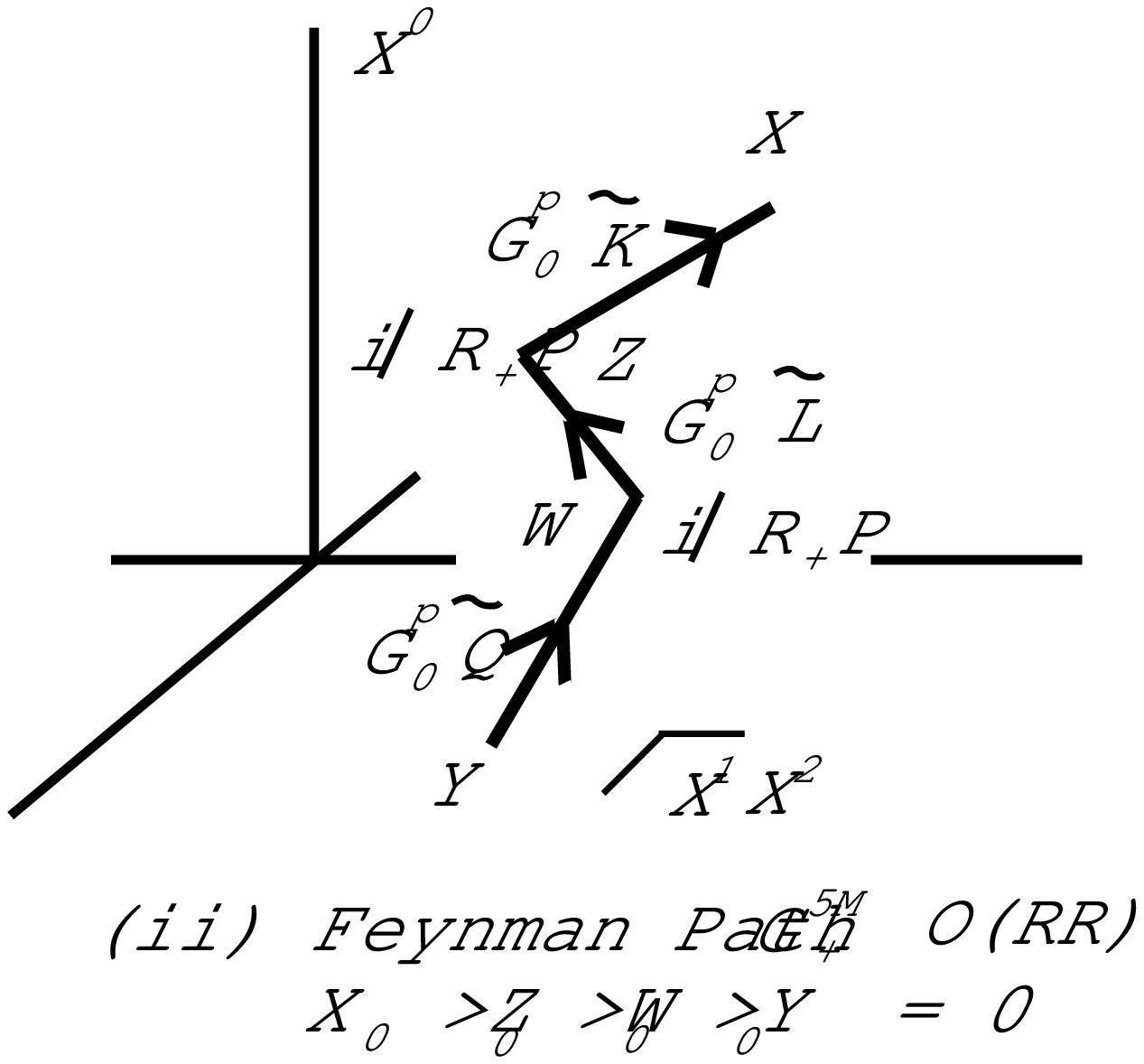}}
   \begin{center}
Fig.9\ 
Non-Abelian Chiral Gauge Theory,$\GfMp$ with Feynman Path , $O(R)$ and $O(RR)$.
   \end{center}
\end{figure}
Using the SC-procedure given in Sec.4, 
the relevant part ($\ep$-tensor $\times \pl R$) is given as
\begin{eqnarray}
\Tr\,i^2\la(x)\GfMp(x,y;t)|_{\ep\pl R}                  \nn
\sim i\int^{X^0}_0dZ^0\int d^2x\tr\,\la(x)\pl_\m R_\n   \nn
\times \int\frac{d^2k}{(2\pi)^2}
\tr\,[-\frac{1}{i}\{ \frac{\pl}{\pl k^\m}\om(k)
-i\frac{\pl E(k)}{\pl k^\m}(X^0-Z^0)\Omp(k)\} \ga_\n P_+\Omp(k)]
\e^{-iE(k)X^0}                                           \nn
\mbox{}                                                  \nn
\sim -\int^{X^0}_0dZ^0\int d^2x(\tr\,\la(x)\pl_\m R_\n)  
\times \int\frac{d^2k}{(2\pi)^2}                         \nn
\left[ (-\frac{M}{2E^4}-\frac{1}{4E^3})\ep_{\la\n}k^\m k^\la
+(\frac{M}{4E^2}+\frac{1}{4E})\ep_\mn
-i(X^0-Z^0)\frac{M}{2E^3}\ep_{\la\n}k^\la k^\m
\right]
\e^{-iE(k)X^0}                                          \nn
\mbox{}                                                 \nn
\sim -\int d^2x(\tr\,\la(x)\pl_\m R_\n)  \times         \nn
\frac{1}{2\pi}\int^\infty_0 dk\cdot k\ i\left[ 
Mt\frac{k^2}{4E^4}+t\frac{k^2}{8E^3}-Mt\frac{1}{4E^2}
-t\frac{1}{4E}+Mt^2\frac{k^2}{8E^3}
\right]\ep_\mn \e^{-E(k)t}                               \nn
\mbox{}                                                 \nn
=-\frac{i}{2\pi}\int d^2x(\tr\,\la(x)\pl_\m R_\n)\ep_\mn\nn
\times \left[
\fourth Mt\,E^3_4(Mt)+\frac{1}{8}Mt\,E^3_3(Mt)
-\fourth Mt\,E^1_2(Mt)-\frac{1}{4}Mt\,E^1_1(Mt)
+\frac{1}{8}(Mt)^2 E^3_3(Mt)
\right]                                                 \nn
\mbox{}                                                 \nn
\ra\q
-\frac{i}{2\pi}(\fourth\times 0+\frac{1}{8}\times 1
-\fourth \times 0-\frac{1}{4}\times 1
+\frac{1}{8}\times 0)
\inttx\,\ep_\mn\tr \la(x)\pl_\m R_\n \nn
=+\frac{i}{16\pi}\inttx\,\ep_\mn\tr \la(x)\pl_\m R_\n
\q (Mt\ra +0)
\com
\label{NA7.5}
\end{eqnarray}
where we use 
$Y^0=0,\ X^0=-it,\ 
aE^3_4(a)\ra 0\ ,\ aE^3_3(a)\ra 1,
\ aE^1_2(a)\ra 0\ ,\ aE^1_1(a)\ra 1,\ a^2E^3_3(a)\ra 0(a\ra\ +0)$
(see App.B). The same result is obtained in the $X^0$-coordinate.
Taking into account the $\GfMm$ contribution, we finally obtain
the non-Abelian anomaly up to the first order of $R$ .
\begin{eqnarray}
\del_\la\,\ln\, J_{NA}=\lim\,
\Tr\,i\la(x)\,(i \GfMp(x,y;t)-i \GfMm(x,y;t)\ )\nn
=\inttx\,\left[+\frac{i}{16\pi} \ep_\mn\tr \la(x)
(\pl_\m R_\n-\pl_\n R_\m)\right]\q 
\pr
\label{NA7b}
\end{eqnarray}
(This turns out to be true even after the second order is taken into account.)

\q The next order is evaluated from 
\begin{eqnarray}
\GfMp|_{RR}
=\int^{X^0}_0dZ^0\int^{Z^0}_0dW_0\int d^2Z\int d^2W \nn
G^p_0(X,Z)i \Rslash(z)P_+G^p_0(Z,W)
                          i \Rslash(w)P_+G^p_0(W,Y)\nn
=\int^{X^0}_0dZ^0\int^{Z^0}_0dW_0\int d^2z\int d^2w
\int\frac{d^2k}{(2\pi)^2}\frac{d^2l}{(2\pi)^2}\frac{d^2q}{(2\pi)^2}\nn
\times (-i)\Omp(k)i \Rslash(z)P_+
(-i)\Omp(l) i \Rslash(w)P_+
(-i)\Omp(q)   \nn
\times \exp\{-i\Ktil(X-Z)-i\Ltil(Z-W)-i\Qtil(W-Y)\}
\pr
\label{NA8}
\end{eqnarray}
See Fig.9(ii). The contribution to the chiral anomaly
($\ep$-tensor $\times RR$) is evaluated as
\begin{eqnarray}
\Tr\,i^2\la(x)\GfMp(x,y;t)|_{\ep RR}         \nn
\sim i^2\half (X^0)^2i^2(-\frac{i}{2})^3
\int d^2x(\tr\,\la(x)R_\m R_\n)              \nn
\times\int\frac{d^2k}{(2\pi)^2}                    
\,\tr\,[P_+\frac{i\kslash}{E(k)}\ga_\n
(2\Omp(k))^2\ga_\m]\e^{-iE(k)X^0}             \nn
\mbox{}                                        \nn 
= \frac{i}{16} (X^0)^2
\int d^2x(\tr\,\la(x)R_\m R_\n)
\int\frac{d^2k}{(2\pi)^2}                    
\times 4\tr\,[P_+\frac{i\kslash}{E(k)}\ga_\n
\frac{M}{E(k)}\Omp(k)\ga_\m]\e^{-iE(k)X^0}             \nn
\mbox{}                                        \nn 
\sim \frac{-i}{4} t^2
\int d^2x(\tr\,\la(x)R_\m R_\n)
\int\frac{d^2k}{(2\pi)^2}
\times 4\tr\,[\frac{i\kslash}{E(k)}\ga_\n
\frac{M}{E(k)}(-\half\frac{i\kslash}{2E(k)}\gago)\ga_\m]\e^{-E(k)t}             \nn
= 0    \com
\label{NA8b}
\end{eqnarray}
where $\tr \kslash\ga_\n\kslash\gago\ga_\m=0$ is used. 
Similarly we have $\Tr\,i^2\la(x)\GfMm(x,y;t)|_{RR}\sim 0$.
Thus we have confirmed the {\it absence} of $\ep_\mn\tr R_\m R_\n$,
which characterize the 2 dim consistent anomaly(non-covariant).
The result (\ref{NA7b}) is true even after the second order
correction and is the {\it half} of the consistent anomaly. 
If we take the symmetric path, instead of the Feynman, in the above,
we indeed have the consistent anomaly.
\begin{eqnarray}
\del_\la\,\ln\, J_{NA}=\half \lim\,
\Tr\,i\la(x)\,(i \GfMp(x,y;t)-i \GfMm(x,y;t)\ )\nn
=\inttx\,\left[+\frac{i}{8\pi} \ep_\mn\tr \la(x)
(\pl_\m R_\n-\pl_\n R_\m)\right]\q 
\pr
\label{NA8c}
\end{eqnarray}
(As for the past literature, see, for example, (13.68)
or (13.128) of \cite{AAR91}, where we should take $a=1$.)

\flushleft{(ii) Covariant Anomaly}

We take the symmetric path and $G^{5M}_{h\pm}$. 
From (\ref{NA3}) and the boundary condition of Symmetric path
solution (\ref{DW13bx}), we obtain
\begin{eqnarray}
\del_\la\,\ln\, J_{NA}
=\half\lim_{M(X^0-Y^0)\ra +0}\Tr\,i\la(x)\,
i G^{5M}_{h+}(x,y;t)\nn
+\half\lim_{M(X^0-Y^0)\ra -0}\Tr\,i\la(x)\,
(-i) G^{5M}_{h-}(x,y;t)+O(\la^2)\pr
\label{NA9}
\end{eqnarray}

The first term is evaluated as
\begin{eqnarray}
G^{5M}_{h+}|_R
=\int^{X^0}_0dZ^0\int d^2Z(G^p_0(X,Z)-G^n_0(X,Z))i \Rslash(z)
(G^p_0(Z,Y)-G^n_0(Z,Y))             \nn
=\int^{X^0}_0dZ^0\int d^2z\int\frac{d^2k}{(2\pi)^2}\frac{d^2l}{(2\pi)^2}\nn
\times (-i)(\Omp(k)\e^{-iE(k)(X^0-Z^0)}
           -\Omm(k)\e^{iE(k)(X^0-Z^0)})
i \Rslash(z)                             \nn
\times (-i)(\Omp(l)\e^{-iE(l)(Z^0-Y^0)}
    -\Omm(l)\e^{iE(l)(Z^0-Y^0)})
\e^{-ik(x-z)-il(z-y)}
\pr
\label{NA9b}
\end{eqnarray}
See Fig.10(i).

\begin{figure}
\centerline{\epsfysize=6cm\epsfbox{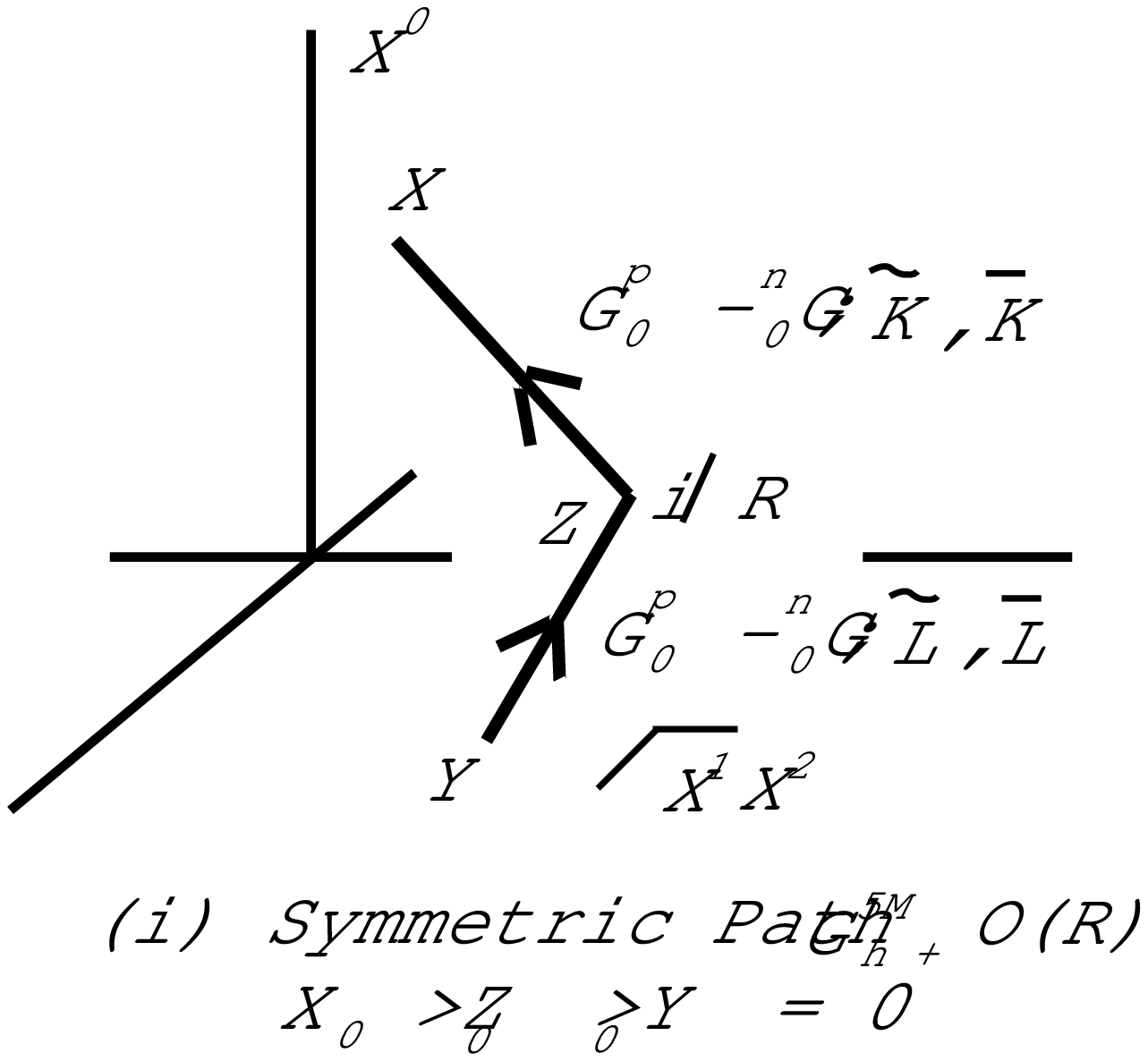}\ \ 
            \epsfysize=6cm\epsfbox{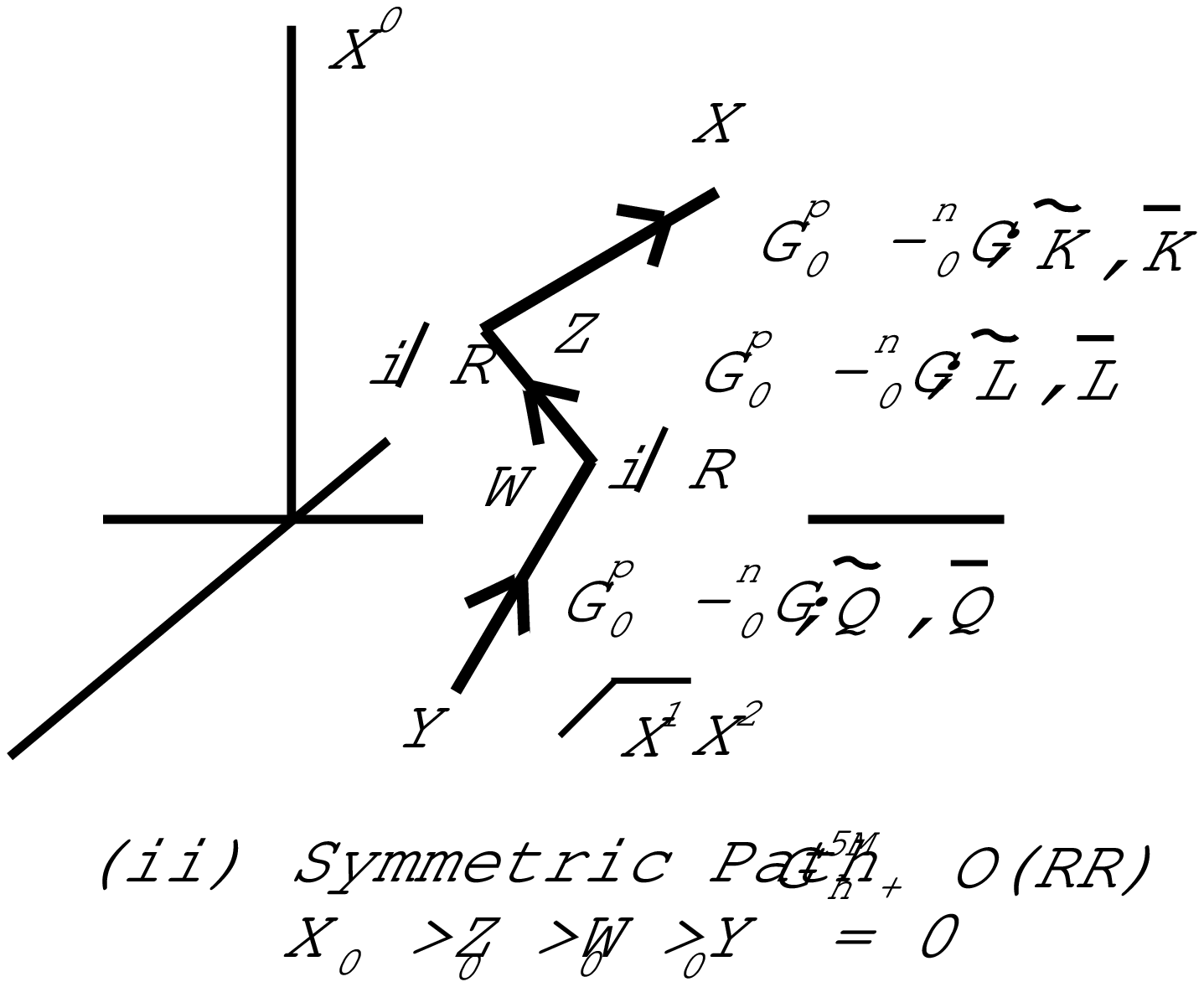}}
   \begin{center}
Fig.10\ 
Non-Abelian Chiral Gauge Theory,$G^{5M}_{h+}$ with Symmetric Path , 
$O(R)$ and $O(RR)$.
   \end{center}
\end{figure}

After the SC-procedure of Sec.4, we are led to
the following one, as the relevant part for the anomaly,
\begin{eqnarray}
\half\Tr\,i^2\la(x)G^{5M}_{h+}(x,y;t)|_{\ep\pl R}
\sim 
\frac{i^2}{2}\int d^2x(\tr\,\la(x)\pl_\m R_\n)\ep_\mn\nn
\times \frac{-i}{4\pi}MX^0 
\{ S^3_3(MX^0)+2S^1_3(MX^0)+S^3_3(MX^0) \}\nn
\ra\q 
+\frac{i}{4\pi}\inttx\,\ep_\mn\tr \la(x)\pl_\m R_\n
\q (MX^0\ra +0)
\com
\label{NA9c}
\end{eqnarray}
Taking into account $G^{5M}_{h-}$, we obtain
\begin{eqnarray}
\del_\la\ln\,J_{NA}=
+\frac{i}{4\pi}\inttx\,\ep_\mn\tr \la(x)(\pl_\m R_\n-\pl_\n R_\m
+O(R^2))
\pr
\label{NA9d}
\end{eqnarray}
The coefficient is two times of the previous case, which is the well-known
relation between 2D consistent and covariant anomalies.

Next order contribution is evaluated from
\begin{eqnarray}
G^{5M}_{h+}|_{RR}
=\int^{X^0}_0dZ^0\int^{Z^0}_0dW^0\int d^2Z\int d^2W
(G^p_0(X,Z)-G^n_0(X,Z))i \Rslash(z)  \nn
(G^p_0(Z,W)-G^n_0(Z,W))i \Rslash(w)
(G^p_0(W,Y)-G^n_0(W,Y))             \nn
=\int^{X^0}_0dZ^0\int^{Z^0}_0dW^0\int d^2z\int d^2w   
\int\frac{d^2k}{(2\pi)^2}\frac{d^2l}{(2\pi)^2}\frac{d^2q}{(2\pi)^2}\nn
\times (-i)(\Omp(k)\e^{-iE(k)(X^0-Z^0)}
           -\Omm(k)\e^{iE(k)(X^0-Z^0)})
i \Rslash(z)                             \nn
\times (-i)(\Omp(l)\e^{-iE(l)(Z^0-W^0)}
           -\Omm(l)\e^{iE(l)(Z^0-W^0)})
i \Rslash(w)                             \nn
\times (-i)(\Omp(q)\e^{-iE(q)(W^0-Y^0)}
           -\Omm(q)\e^{iE(q)(W^0-Y^0)})
\e^{-ik(x-z)-il(z-w)-iq(w-y)}
\pr
\label{NA10}
\end{eqnarray}
See Fig.10(ii). After the use of SC-procedure, we obtain
\begin{eqnarray}
\Tr\,\la(x)G^{5M}_{h+}|_{\ep RR}
\sim
(-i)\int^{X^0}_0dZ^0\int^{Z^0}_0dW^0\int d^2x   
\int\frac{d^2k}{(2\pi)^2}\tr\,[\la(x)R_\m(x)R_\n(x)]\nn
\times \{ \om(k)(-2i)\sin E(k)(X^0-Z^0)
           +\gago\cos E(k)(X^0-Z^0) \}\ga_\m          \nn
\times \{ \om(k)(-2i)\sin E(k)(Z^0-W^0)
           +\gago\cos E(k)(Z^0-W^0) \}\ga_\n          \nn
\times \{ \om(k)(-2i)\sin E(k)(W^0-Y^0)
           +\gago\cos E(k)(W^0-Y^0)\}                 \nn
\mbox{}\nn		   
\sim		   
(-i)\inttx\tr\,[\la(x)R_\m(x)R_\n(x)]\nn
\times\int\frac{d^2k}{(2\pi)^2}[(\tr\,\gago\ga_\m\gago\ga_\n\gago)\fourth
\{\half (X^0)^2\cos EX^0+\frac{3X^0}{2E}\sin EX^0\}\nn
+(\tr\,\gago\ga_\n\ga_\m)(-2i)^2\nn
\times\{
\frac{M^2}{4E^2}(-\cos E(X^0-Z^0)\sin EW^0-\sin E(X^0-Z^0)\cos EW^0)
\sin E(Z^0-W^0)        \nn
+\frac{M^2-k^2}{4E^2}\sin E(X^0-Z^0)\cos E(Z^0-W^0)\sin EW^0\}]\nn
\mbox{}\nn
=(-i)\inttx\tr\,[\la(x)R_\m(x)R_\n(x)]\nn
\times 2i\ep_\mn [\ 
\fourth \{ \half\frac{(MX^0)^2}{2\pi}C^1_0(MX^0)
+\frac{3}{2}\frac{MX^0}{2\pi}S^1_1(MX^0)\}       \nn
+(-2i)^2\fourth \{ -\frac{MX^0}{8\pi}S^1_3(MX^0)  
+\frac{(MX^0)^2}{16\pi}( C^1_2(MX^0)-C^3_2(MX^0) ) \}\ ]    \nn
\ra \inttx\tr\,[\la(x)R_\m(x)R_\n(x)]\times (-i)\times 2i\ep_\mn
\times (+\frac{1}{4\pi})\pr		   
\label{NA11}
\end{eqnarray}
Adding $G^{5M}_{h-}$ contribution, the above result gives indeed 
$O(R^2)$ part of the covariant anomaly.
\begin{eqnarray}
\del_\la\ln\,J_{NA}=
+\frac{i}{4\pi}\inttx\,\ep_\mn\tr \la(x)(\pl_\m R_\n-\pl_\n R_\m
+i[R_\m,R_\n])
\com
\label{NA12}
\end{eqnarray}
which agrees with (13.72) of \cite{AAR91}.

\q We have checked if we take the Feynman path, even with 
$G^{5M}_{h\pm}$, the result leads to the consistent anomaly.

\section{Massive Fermion}
So far we have examined the regularization only for 
the massless fermion. 
Let us consider here the case where the (4 dim) fermion has
a small mass $m$. QED is taken as the example.

\begin{eqnarray}
\Lcal=\psibar\Dhat_m\psi\com\q
\Dhat_m=\Dhat-m\com\q
{\Dvec_m}^\dag=\Dvec_m\pr
\label{MF1}
\end{eqnarray}
where $\Dhat=i\ga_\m(\pl_\m+i\e A_\m)$ has been introduced in (\ref{DW1}).
In the same way as eq.(\ref{DW2}), the effective action is given by
\begin{eqnarray}
\ln\,Z_m[A]=-\Tr\int_0^\infty\frac{\e^{-t\Dhat_m}}{t}dt+\mbox{const}\nn
=-\int_0^\infty\frac{dt}{t}
\Tr\,[\half(1+i\ga_5)\Tr \e^{+it\gago (\Dhat-i\gago m)}+
\half(1-i\ga_5)\Tr \e^{-it\gago(\Dhat+i\gago m)}]+\mbox{const}.
\label{MF2}
\end{eqnarray}
The corresponding heat-kernels satisfy the following
1+4 dim massive Dirac equation.

\begin{eqnarray}
\ln\,Z_m=-\lim_{M\ra 0}\int_0^{\infty}\frac{dt}{t}
\half(1-i\frac{\pl}{t\,\pl M})\Tr G^{5M}_{m+}(x,y;t)  \nn
-\lim_{M'\ra 0}\int_0^{\infty}\frac{dt}{t}
\half(1-i\frac{\pl}{t\,\pl M'})\Tr G^{5M'}_{m-}(x,y;t)\com\nn
\mbox{where}\nn
G^{5M}_{m+}(x,y;t)\equiv <x|\exp\{+it\gago(\Dhat-i\gago m+iM)\}|y>\com\nn
(i\dbslash-M)G^{5M}_{m+}=(i\e \Aslash-\gago m)G^{5M}_{m+}\com\q
(X^a)=(-it,x^\m)\com                    \nn
G^{5M'}_{m-}(x,y;t)\equiv <x|\exp\{-it\gago(\Dhat+i\gago m+iM')\}|y>\com\nn
(i\dbslash-M')G^{5M'}_{m-} =(i\e \Aslash+\gago m) G^{5M'}_{m-}\com\q
(X^a)=(+it,x^\m)   \pr
\label{MF3}
\end{eqnarray}
\footnote{
If we write as
$i\e \Aslash\mp\gago m=-e\Acalbslash\ ,\ 
({\cal A}^a)=(\pm \frac{m}{e},A_\m)$, 
the theory looks like the 1+4 dim QED in a {\it temporal} gauge 
(${\cal A}^0=\pm \frac{m}{e}$). 
In the lattice approach, however, such an extended standpoint does not 
seem to successfully work at present\cite{Sha93}.
}


\q In the lattice approach, the superiority of the domain wall fermion
(to the ordinary Wilson fermion) is no need to fine-tune the
hopping parameter which suffers from the renormalization due to
the gauge interaction. 
In addition to the wave-function and coupling renormalizations,
the mass renormalization is one crucial test of the superiority.
We can calculate such an effect using the above formulae. The full 
consideration can be done only after 
the fermion and
the gauge fields are treated on the equal footing.
It has recently been performed in \cite{SI99aei}. 

\section{Conclusion and Discussion}
Inspired by the recent progress of the chiral fermion on lattice,
we have presented a new regularization, for 
the continuum Euclidean field theories, which is
based on the domain wall
configuration in one dimension higher (Minkowski) space. 
We have verified the analogous aspect to the lattice case,
such as the domain wall structure, the condition on the
1+4 dim Dirac mass (the regularization parameter $M$),
the overlap Dirac operator, etc. Applying the proposed
regularization to some models (4 dim QED, 2 dim chiral gauge theory), 
the known anomalies
are correctly reproduced. It shows the present regularization
correctly works.

\q Comparing Fig.2 (or 3) with Fig.4, we can imagine that the choice
of (anti-)Feynman path solution perturbatively defines the chiral version
of the original theory, for example, the chiral gauge theory. As far as
anomaly calculation is concerned, it holds true. In order to show the
statement definitely, we must clarify the following things. The "ordinary"
chiral symmetry appears only in the limit :\ $\frac{|k^\m|}{M}\ra +0$, as
shown in (\ref{DW9by}). But this limit can not be taken because it "freezes"
the dynamics and anomalies do {\it not} appear.  It seems we must introduce
some new "softened" version of the chiral symmetry which keeps the dynamics.
One standpoint taken in this paper is to replace 
$|k^\m|/M\ \ll 1$\ by $|k^\m|/M\ \leq 1$\ in (\ref{DW9by}).
It breaks the "ordinary" chiral symmetry. It could, however, be
possible that this replacement can avoid the breaking
by changing (generalizing) the "ordinary" chiral symmetry. 
In this case, the new chiral Lagrangian has infinitely many higher-derivative
terms. We should explain this new "deformed" Lagrangian from some generalized
chiral symmetry. This situation looks similar to what
L\"{u}sher\cite{Lus98} did for the chiral lattice.

\q The present regularization is characterized by the introduction of the
fifth coordinate $t$, and the regularization mass $M$. The 
limit (\ref{DW14}) is quite impressive as a new regularization. Because we are
, in this article, mainly concerned with anomalies, the analysis need not
to touch upon the $t$-integral. (We examine only the variation of the partition
function under the chiral transformation, or the measure change. We do not
examine the partition function itself. ) For the evaluation of the partition
function, we know $t$-integral is usually regularized as
\begin{eqnarray}
\ep \leq t \leq T\q (\frac{1}{\ep}\geq \frac{1}{t} \geq \frac{1}{T})\com\q
\frac{T}{\ep}\gg 1\com
\label{Conc1}
\end{eqnarray}
where another two regularization parameters $\ep$(ultraviolet)
and $T$(infrared) appear. The consistency with (\ref{DW14}) requires
\begin{eqnarray}
M\ll \frac{1}{\ep}\pr
\label{Conc2}
\end{eqnarray}
The full treatment of the partition function has been consistently 
done in a recent work\cite{SI99aei}, where the renormalization
is formulated with the care for the quantum effect of
both fermions and gauge bosns.

\q 
We stress
characteristic points of the present
domain wall regularization.
\begin{itemize}
\item
The higher dimensional Dirac field mass $M$ is the only
regularization parameter used in this paper. It should satisfy
the condition (\ref{DW14}) in order to do the following role:\ 
1)\ controlling the chirality, 
2)\ forming the domain wall configuration (dimensional reduction), 
3)\ Ultraviolet and infrared regularization for the momentum integral.
\item
There appears different regularizations depending on the solutions
of the higher dim Dirac equation.
The Feynman path is (practically) appropriate for the analysis of
chiral properties, although overall factors of anomalies deviate
from the right ones. 
The calculation is most simple.  
Whereas the symmetric path gives the proper solution of the Dirac
equation and the regularization with the path gives the
correct anomaly including the coefficient. 
Especially 
the covariant anomaly
is obtained only by the choice of the symmetric path and the hermitian vertex.
Other choices essentially lead to
the consistent anomaly.
\item
The characteristic functions appearing in the present regularization,
such as the incomplete gamma function, indicate the advance
over other ordinary regularizations,
say, the dimensional one where only the gamma function appears.
\end{itemize}

\q The chiral problem itself does not depend on the interaction.
It looks a kinematical problem in the quantization of fields.
How do we treat the different propagations of {\it free} solutions
depending on the
boundary conditions (with respect to the Wick-rotated time) 
is crucial to the problem.
In the standpoint of the operator formalism 
(the Fock-space formalism) 
it corresponds to how to treat the "delicate" structure
(due to the ambiguity of the fermion mass sign) 
of the vacuum of the free fermion theory.
The present paper insists the following prescription:\ 
First we go to 1+4 dim Minkowski space by the Wick-rotation
of the inverse temperature $t$, and take the "directed"
solution as in Sect.3. 
The anomaly phenomena concretely reveal  
the chiral problem. The proposed prescription passes the anomaly test.

\q The recent progress in the lattice formalism, using the Neuberger's
overlap Dirac operator\cite{Neub98} and L{\"{u}}scher's chiral symmetry
on lattice\cite{Lus98},  
has revealed the importance of the condition
on $M$, where the anomaly coefficient varies depending on the
different conditions on $M$.
$M$ is bounded by the Wilson parameter
in some ways\cite{KY99,KF99,Ada98,Suzu98}.
It shows whether the zero mode (surface state) is correctly picked up
in the regularization crucially depends on the choice of $M$. 
Note that there is not the Wilson term in the present formalism. Even such case
we have the similar condition (\ref{DW14}) which produces the correct
coefficient. 
This strongly implies the present
regularization effectively does the same thing as the Wilson term does in the
lattice formalism.


\vspace{2cm}
{\large Appendix A. Notations}
\vspace{0.5cm}

In the analysis of the chiral property, much care should be
paid to $\pm 1$ and $\pm i$. The physical interpretation
of the final result heavily relies on such delicacy. 
In this circumstance, we decide to present the present
notation in this appendix.
We adopt the convention of Ref.\cite{NN95} for the gamma matrices
($\ga_\m$) and the metric in 2 and 4 dim Euclidean space and 
for those ($\Ga_a$) in 5 dim Minkowski space. 

\flushleft{(ia) 2 dim Euclidean}:$\m=\mbox{1,2}$.
\begin{eqnarray}
\ga_1=\left( \matrix{0, & 1 \cr
                     1, & 0     } \right)
=\si_1\ ,\ 
\ga_2=\left( \matrix{0, & i \cr
                     -i, & 0     } \right)
=-\si_2\ ,\ 
\gago=i\ga_1\ga_2
=\left( \matrix{1, & 0 \cr
                0, & -1     } \right)
=\si_3\ ,\nn 
\gago^\dag=\gago\ ,\ 
\Tr\,\ga_\m\ga_\n\gago=-2i\ep_\mn\ ,\ \ep_{12}=1\ ,\nn
\{\ga_\m,\ga_\n\}=2\del_\mn\ ,\ \ga_\m^\dag=\ga_\m\ ,\ 
(\del_\mn)=\mbox{diag}(1,1)\com \nn
\psibar=\psi^\dag\ga_2\com 
\label{AppA.1}
\end{eqnarray}
where $\si_i$($i=$1,2,3) are Pauli matrices.

\flushleft{(ib) 1+2 dim Minkowski}:$a=0,1,2;\ \m=1,2$.
\begin{eqnarray}
\{\Ga_a,\Ga_b\}=2\eta_{ab}\ ,\ 
(\eta_{ab})=\mbox{diag}(1,-1,-1)\ ,\ 
\Ga_0=\Ga^0=\gago\ ,\  \Ga_\m=-\Ga^\m=-i\ga_\m\ ,\nn 
{\Ga_0}^\dag=\Ga_0\ ,\ {\Ga_\m}^\dag=-\Ga_\m\ ,\ 
\psibar=\psi^\dag\Ga_0=\psi^\dag\gago
\label{AppA.2}
\end{eqnarray}

\flushleft{(iia) 4 dim Euclidean}:$\m=\mbox{1,2,3,4}$.
\begin{eqnarray}
\ga_\m=\left( \matrix{0,           & \si_\m \cr
                      \si_\m^\dag, & 0         } \right)\ ,\ 
\{\ga_\m,\ga_\n\}=2\del_\mn\com\q \ga_\m^\dag=\ga_\m\com\q
(\del_\mn)=\mbox{diag}(1,1,1,1)\com    \nn
\gago=\ga_1\ga_2\ga_3\ga_4=\left( \matrix{1_{2\times 2}, & 0 \cr
                                       0,   & -1_{2\times 2}} \right)\com\q
\gago^\dag=\gago\ ,\ (\gago)^2=1\com\nn
\Tr\ga_\m\ga_\n\ga_\la\ga_\si\gago=4\ep_{\mn\ls}\ ,\ \ep_{1234}=1\ ,\ \nn
\mbox{chiral projection operator:}\q P_\pm\equiv
\half (1_{4\times 4}\pm\gago)\com\nn
\psibar=\psi^\dag\ga_4\com
\label{AppA.3}
\end{eqnarray}
where $\si_i$($i=$1,2,3) are Pauli matrices defined in (ia), and 
$\si_4=\left(\matrix{i,& 0\cr 0,& i}\right)$.

\flushleft{(iib) 1+4 dim Minkowski}:$a=0,1,2,3,4;\ \m=1,2,3,4$.

\begin{eqnarray}
\{\Ga_a,\Ga_b\}=2\eta_{ab}\ ,\ 
(\eta_{ab})=\mbox{diag}(1,-1,-1,-1,-1)\ ,\nn 
\Ga_0=\Ga^0=\gago\ ,\  \Ga_\m=-\Ga^\m=-i\ga_\m\ , 
\psibar=\psi^\dag\Ga_0=\psi^\dag\gago
\label{AppA.4}
\end{eqnarray}

\vspace{2cm}
{\large Appendix B. Integral Formulae}
\vspace{0.5cm}

In Sec.5 of the text, we explain the momentum integral. We list
here some useful integral formulae. 
We define $E_n^r(a),\ S_n^r(a),\ C_n^r(a)$ as
\begin{eqnarray}
a>0\com\q n=0,1,2,3,\cdots\ \mbox{(non-negative integer)}\com\nn 
r=1,3,5,\cdots\ \mbox{(positive odd integer)}\com\nn
E_n^r(a)\equiv \int^\infty_0 dx\frac{x^r}{(\sqxx)^n}
\e^{-a\sqxx}\com\nn
S_n^r(a)\equiv \int^\infty_0 dx\frac{x^r}{(\sqxx)^n}
\sin (a\sqxx)\com\nn
C_n^r(a)\equiv \int^\infty_0 dx\frac{x^r}{(\sqxx)^n}
\cos (a\sqxx)\com\nn
E_n^r(-ia)=C_n^r(a)+i\,S_n^r(a).
\label{Int.1}
\end{eqnarray}
Some exact expressions are
\begin{eqnarray}
E_3^3(a)=
\int^\infty_0 dx\frac{x^3}{(\sqxx)^3}\e^{-a\sqxx}
=\frac{1}{a}\Ga(1,a)-a\Ga(-1,a)               \nn
=\frac{1}{a}(1-2a+O(a^2,a^2\ln a))\com\nn
E_3^3(a)+2E_3^1(a)=
\int^\infty_0 dx\frac{x(x^2+2)}{(\sqxx)^3}\e^{-a\sqxx}
=\frac{1}{a}\Ga(1,a)+a\Ga(-1,a)               \nn
=\frac{1}{a}(1+0\times a+O(a^2))\com\nn
E_2^1(a)=
\int^\infty_0 dx\frac{x}{x^2+1}\e^{-a\sqxx}=-\mbox{Ei}(-a)=\Ga(0,a) \nn
=-\ln a-\ga+O(a)\com\nn
S_5^5(a)+2S_5^3(a)
=\int^\infty_0 dx\frac{x^3(x^2+2)}{(\sqxx)^5}\sin (a\sqxx)   \nn
=\frac{1}{720 a}\{
-360 a^2+220 a^4-120 a^4\ga+720 \cos a    \nn
+3a^6\,\mbox{}_2F_3(1,1;2,3,7/2;-a^2/4)-120 a^4\ln a\}   \nn
=\frac{1}{a}(1+O(a^2))             \com\nn
C_4^3(a)+2C_4^1(a)
=\int^\infty_0 dx\frac{x(x^2+2)}{(x^2+1)^2}\cos (a\sqxx)   \nn
=\half-\frac{3}{4}a^2+\frac{a^2}{2}\ga-\mbox{Ci}(a)
-\frac{a^4}{48}\,\mbox{}_2F_3(1,1;2,5/2,3;-a^2/4)
+\frac{a^2}{2}\ln\,a        \nn
=\half-\ga-\ln\,a+O(a^2)      \com
\label{Int.2}
\end{eqnarray}
where $\ga=0.57721\cdots$ is the Euler constant,
$\Ga(*,*)$, Ei($*$), Ci($*$) and $\mbox{}_pF_q(*;*;*)$
are the incomplete gamma function, the exponential integral
function, the cosine integral function and the  
(Pochhammer's) generalized hypergeometric function respectively.
They are defined as
\begin{eqnarray}
\Ga(z,p)=\int^\infty_p\e^{-t}t^{z-1}dt,\q \Ga(z)=\int^\infty_0 \e^{-t}t^{z-1} dt 
=\Ga(z,p=0),\q \mbox{Re}\,z>0\q ;\nn
\Ga(1-z,p)\Ga(z)=p^{1-z}\int^\infty_0
\frac{\e^{-(p+t)}t^{z-1}}{p+t} dt\com\q \mbox{Re}\,z>0\q ;\nn
\mbox{Ei}(-x)=-\int^\infty_x\frac{\e^{-t}}{t} dt
=-\Ga(0,x)\com\q x>0\q ;\nn
\mbox{Ci}(x)=-\int^\infty_x\frac{\cos t}{t} dt
=\ga+\ln\,x+\int^x_0\frac{\cos\,t-1}{t}dt
\com\q x>0\q ;\nn
\mbox{}_pF_q(\al_1,\al_2,\cdots,\al_p;\be_1,\be_2,\cdots,\be_q;z)\nn
=\sum^\infty_{n=0}
\frac{(\al_1)_n\cdots (\al_p)_n}{(\be_1)_n\cdots (\be_q)_n}
\frac{z^n}{n!}\com\q (\al)_n=\frac{\Ga(\al+n)}{\Ga(\al)}\ ,\ (\al)_0=1
\ ,\ |z|<1\ ;
\label{Int.3}
\end{eqnarray}
where $\Ga(z)$
is the (Euler) gamma function. Some other formulae useful for
the weak expansion are 
\begin{eqnarray}
\mbox{Re}\,z>0\ ,\ \Ga(z,p)=\Ga(z)-\ga(z,p)\ ,\ 
\ga(z,p)=\int^p_0\e^{-t}t^{z-1}dt
=\sum^\infty_{n=0}\frac{(-1)^np^{z+n}}{n!(z+n)},\nn
\mbox{Ei}(-x)=\ln\,x+\ga+O(x)\ ,\  1\gg x>0\ .
\label{Int.3b}
\end{eqnarray}
The following asymptotic expressions for $a\ra +0$  are 
used in the text.
\begin{eqnarray}
aE_1^1(a)\ra 1\ ,\ aE_2^1(a)\ra 0\ ,\ a^2E_2^3(a)\ra 1\ ,\ 
aE_3^3(a)\ra 1\ ,\ aE_3^1(a)\ra 0\ ,           \nn
a^2E_3^3(a)\ra 0\ ,\ aE_4^1(a)\ra 0\ ,\ aE_4^3(a)\ra 0\ ,\ 
a^2E_4^3(a)\ra 0\ ,\ a^2E_4^5(a)\ra 1\ ,       \nn
a(S_3^3(a)+2S_3^1(a))\ra 1\ ,\ 
a(C_3^3(a)+2C_3^1(a))\ra 0\ ,  \nn
a(C_4^3(a)+2C_4^1(a))\ra 0\ ,\ 
a(S_5^5(a)+2S_5^3(a))\ra 1\ ,
\label{Int.4}
\end{eqnarray}

\vspace{2cm}
{\large Appendix C. Projective Properties of $\Om_\pm(k)$}
\vspace{0.5cm}

Here we list some useful relations of 
$\Om_\pm(k)$, (\ref{DW8}), which correspond to free solutions
of 1+4 dim Dirac equation.
$E(k)=\sqrt{k^2+M^2},(\Ktil^a)=(E(k),K^\m=-k^\m),
(\Kbar^a)= (E(k),-K^\m=k^\m)$. 
$k^\m$ is the momentum in the 4 dim Euclidean space.
$\Ktil$ and $\Kbar$ are on-shell
momenta, of 1+4 dim Dirac equation, 
($\Ktil^2=\Kbar^2=M^2$), which correspond to the positive
and negative energy states respectively.
\begin{eqnarray}
\Om_+(k)\equiv\frac{M+\Ktilbslash}{2E(k)}=\om(k)+\frac{\gago}{2}\com\q
\om(k)\equiv \frac{M+i\kslash}{2E(k)}\com\nn
\Om_-(k)\equiv\frac{M-\Kbarbslash}{2E(k)}=\om(k)-\frac{\gago}{2}
=\Om_+(k)-\gago\com\q\nn
(\frac{\gago}{2})^2=
\om(k)\om^\dag(k)=\om^\dag(k)\om(k)=\fourth I_{4\times 4}\com\nn
\om(k)^\dag=\om(-k)= \frac{M-i\kslash}{2E(k)}\com\nn
\Om_+(k)^\dag=\Omp(-k)=\frac{M-i\kslash+E\gago}{2E(k)}\com\q
\Om_-(k)^\dag=\Omm(-k)=\frac{M-i\kslash-E\gago}{2E(k)}\com
\label{PR1}
\end{eqnarray}
In the following we will omit the momentum ($k$) dependence
of $\Om_\pm(k),\om(k)$
unless we would like to stress
the momentum coordinate.
\begin{eqnarray}
\Om_+-(\Om_+)^\dag=\Om_--(\Om_-)^\dag=\om-\om^\dag=
\frac{i}{E}\kslash\com\nn
\Om_++(\Om_+)^\dag=\frac{M}{E}+\gago\ ,\ 
\Om_-+(\Om_-)^\dag=\frac{M}{E}-\gago\ ,\ 
\om+\om^\dag=\frac{M}{E}\ ,\ \nn
\mbox{(1A)}\q 
\Om_++(\Om_-)^\dag=\frac{M}{E}\com\q 
\mbox{(1B)}\q
\Om_-+(\Om_+)^\dag=\frac{M}{E}\com\nn
\mbox{(2A)}\q 
\Om_+-(\Om_-)^\dag=\frac{i}{E}\kslash+\gago\com\q
\mbox{(2A')}\q
(\frac{i}{E}\kslash+\gago)^2=(\frac{M}{E})^2\com\nn
\mbox{(2B)}\q
\Om_--(\Om_+)^\dag=\frac{i}{E}\kslash-\gago\com\q
\mbox{(2B')}\q
(\frac{i}{E}\kslash-\gago)^2=(\frac{M}{E})^2\com\nn
\Om_++\Om_-=2\om\com\q \Om_+-\Om_-=\gago\com\nn
\mbox{(3)}\q 
\Omp(\Omm)^\dag=(\Omp)^\dag\Omm=0\com\nn
\Omp(\Omp)^\dag=\gago\Omp(-k)\com\q 
\Omm(\Omm)^\dag=-\gago\Omm(-k)\com\nn
\mbox{(4A)}\q 
\Omp\Omp=\frac{M}{E}\Omp\com\q 
\mbox{(4B)}\q
\Omm\Omm=\frac{M}{E}\Omm\com\nn
\Omp\gago\Omp=\Omp\com\q \Omm\gago\Omm=-\Omm\com\nn
\Omp\Omm=\frac{M}{E}\Omp(k)-\gago\Omp(-k)
=-\frac{k^2}{2E^2}+\frac{i}{2E}(\frac{M}{E}+\gago)\kslash\ ,\nn
\Omm\Omp=\frac{M}{E}\Omm(k)+\gago\Omm(-k)
=-\frac{k^2}{2E^2}+\frac{i}{2E}(\frac{M}{E}-\gago)\kslash\ ,\nn
\left[ \gago\ ,\ \om \right] =\frac{i}{E}\gago\kslash\com\q
\left[ \ga_\m\ ,\ \om \right] =\frac{k^\n}{E}\si_\mn\com
\si_\mn\equiv \frac{i}{2}[\ga_\m\ ,\ \ga_\n ]\com\q
(\si_\mn)^\dag=\si_\mn\ ,\nn
\om\gago=\gago\om^\dag\com\q
\Omp\gago=\gago (\Omp)^\dag\com\q \Omm\gago=\gago (\Omm)^\dag\com\nn
\left[ \ga_\m\ ,\ \Omp \right] =\frac{k^\n}{E}\si_\mn+\ga_\m\gago\com\q
\left[ \ga_\m\ ,\ \Omm \right] =\frac{k^\n}{E}\si_\mn-\ga_\m\gago\com\nn
\{ \ga_\m ,\Om_\pm\}=\{ \ga_\m ,\om\}=\frac{M\ga_\m+ik^\m}{E}\com
\label{PR2}
\end{eqnarray}
The numbered equations above show
projective property between $\Omp$ and $\Omm^\dag$ and between
$\Omm$ and $\Omp^\dag$.

\vs 1
\begin{flushleft}
{\bf Acknowledgment}
\end{flushleft}
The author thanks K.Fujikawa for discussions at intermidiate stages.
He also thanks
N.Ikeda for fruitful discussions at some points.
Finally the author thanks N.Nakanishi and M.Abe for pointing out
something about the general solution of the Dirac equation.

\vs 1



\begin{thebibliography}{99}
\bibitem{Wil75} 
K.G.Wilson, {New phenomena in sub-nuclear physics (Erice,1975),
ed. A.Zichichi (Plenum, New York,1977),Part A,69}
\bibitem{Creu83} 
M.Creutz,{"Quarks,gluons and lattices", Cambridge Univ. Press,
Cambridge, 1983}
\bibitem{NN81} 
H.Nielsen and M.Ninomiya, {\NP {\bf B185}(1981)20;
                               {\bf B193}(1981)173}
\bibitem{Kap92} 
D.B.Kaplan,{\PL{\bf B288}(1992)342}
\bibitem{Jan92} 
K.Jansen,{\PL{\bf B288}(1992)348}
\bibitem{NN94} 
R.Narayanan and H.Neuberger,{\NP{\bf B412}(1994)574;\PRL{\bf 71}(1993)3251}
\bibitem{NN95} 
R.Narayanan and H.Neuberger,{\NP{\bf B443}(1995)305}
\bibitem{DS95PL} 
Randjbar-Daemi and J.Strathdee,{\PL{\bf B348}(1995)543}
\bibitem{DS95NP} 
Randjbar-Daemi and J.Strathdee,{\NP{\bf B443}(1995)386}
\bibitem{DS97PL} 
Randjbar-Daemi and J.Strathdee,{\PL{\bf B402}(1997)134}
\bibitem{Sha93} 
Y.Shamir,{\NP{\bf B406}(1993)90}
\bibitem{CH94} 
M.Creutz and J.Horv{\'a}th,{\PR{\bf D50}(1994)2297}
\bibitem{Creu95} 
M.Creutz,{\NP B(Proc.Suppl.){\bf 42}(1995)56}
\bibitem{FS95} 
V.Furman and Y.Shamir,{\NP{\bf B439}(1995)54}
\bibitem{Vra98} 
P.M.Vranas,{\PR{\bf D57}(1998)1415}
\bibitem{Col99}
P.Vranas,P.Chen,N.Christ,G.Fleming,A.Kaehler,C.Malureanu,
R.Mawhinney,G.Siegert,C.Sui,L.Wu, and Y.Zhestkov,
{hep-lat/9903024,"Dynamical lattice QCD thermodynamics and
the U(1)$_A$ symmetry with domain wall fermions"}
\bibitem{Neub98} 
H.Neuberger,{\PL{\bf B417}(1998)141}
\bibitem{GW82}  
P.H.Ginsparg and K.G.Wilson, {\PR{\bf D25}(1982)2649}
\bibitem{Lus98}  
M.L{\"{u}}scher, {\PL{\bf B428}(1998)342}
\bibitem{SI98}   
S.Ichinose,{BNL-preprint, Univ.of Shizuoka preprint,
US-98-09, hep-th/9811094, to be published in Phys.Rev.D,
"Temperature in Fermion Systems and
the Chiral Fermion Determinant"}
\bibitem{II98}   
S.Ichinose and N.Ikeda,{
\JMP{\bf 40}(1999)2259
}
\bibitem{Sch51} 
J.Schwinger,{\PR{\bf 82},664(1951)}
\bibitem{KF79} 
K.Fujikawa,{\PRL{\bf 42}(1979)1195;{\bf 44}(1980)1733;
\PR{\bf D21}(1980)2848; {\bf D22}(E)(1980)1499}
\bibitem{II96} 
S.Ichinose and N.Ikeda,{\PR{\bf D53}(1996)5932}
\bibitem{KF93} 
K.Fujikawa,{\PR{\bf D48}(1993)3922}
\bibitem{BD} 
J.D.Bjorken and S.D.Drell,{{\it Relativistic Quantum Mechanics}
(McGraw-Hill,New York,1964);{\it Relativistic Quantum Fields}
(ibid.,1965)}
\bibitem{AN92} 
M.Abe and N.Nakanishi,{\PTP {\bf 88}(1992)975}
\bibitem{BS97PR} 
T.Blum and A.Soni,{\PR{\bf D56}(1997)174;
\PRL{\bf 19}(1997)3595}
\bibitem{Leut85} 
H.Leutwyler, {\PL {\bf B152}(1985)78}
\bibitem{DAMTP9687} 
S.Ichinose and N.Ikeda,{DAMTP/96-87, hep-th/9610136,
"Gauge Symmetry of the Heat Kernel and Anomaly Formulae".}
\bibitem{Neub98b} 
H.Neuberger,hep-lat/9802033,"Geometrical aspects of chiral anomalies
in the overlap"
\bibitem{AAR91} 
E.Abdalla, M.C.B.Abdalla, and K.D.Rothe,{{\it 2 Dimensional Quantum Field Theory}
(World Scientific,Singapore,1991)}
\bibitem{SI99aei} 
S.Ichinose,AEI-1999-35,hep-th/9911079,
"Renormalization using Domain Wall Regularization"
\bibitem{KY99} 
Y.Kikukawa and A.Yamada,{\PL {\bf B448}(1999)265}
\bibitem{KF99} 
K.Fujikawa,{\NP {\bf B546}(1999)480}
\bibitem{Ada98} 
D.H.Adams,hep-lat/9812003,"Overlap topological charge and
axial anomaly for lattice fermions with overlap-Dirac operator"
\bibitem{Suzu98} 
H.Suzuki,IU-MSTP/31,hep-th/9812019,
"Simple Evaluation of Chiral Jacobian with Overlap Dirac Operator"
\end{thebibliography}
\end{document}